\documentclass[amssymb,aps,pra,twocolumn,footinbib]{revtex4-2}%longbibliography

\usepackage{subfigure}
\usepackage{enumitem}
\usepackage[colorlinks=true,linkcolor=blue,urlcolor=blue,citecolor=blue]{hyperref}
\usepackage{graphicx}% Include figure files
\usepackage{dcolumn}% Align table columns on decimal point
\usepackage{bm}% bold math
\usepackage{amsmath}
\usepackage{amsfonts}
\usepackage{amssymb}

\usepackage{tabularx}
\usepackage[overload]{empheq}
\usepackage{eufrak}
\usepackage{siunitx}

\newcommand{\PathToFigures}{} % Put here the path to the folder with figures!

\DeclareSIUnit{\angstrom}{\textup{\AA}}

\begin{document}

\title{Surface quantum critical phenomena in disordered Dirac semimetals}

\author{Eric Brillaux,$^1$ Andrei A. Fedorenko,$^1$\ and Ilya A. Gruzberg$^2$ }

\affiliation{\mbox{$^1$ Univ Lyon, ENS de Lyon, CNRS, Laboratoire de Physique, F-69342 Lyon, France}\\
\mbox{$^2$ Ohio State University, Department of Physics, 191 West Woodruff Ave, Columbus OH, 43210, USA}}
\date{May 7, 2024}

%%%%%%%%%%%%%%%%%%%%%%%%%%%%%%%%%%%%%%%
%%%%%%%%%%%%%%%%%%%%%%%%%%%%%%%%%%%%%%%
\begin{abstract}

We study a non-Anderson disorder driven quantum phase transition in a semi-infinite Dirac semimetal with a flat boundary. The conformally invariant boundary conditions, which include those that are time-reversal invariant, lead to nodal-like surface states on the boundary. In this case the boundary becomes metallic at a critical disorder that is weaker than that for the semimetal-diffusive metal transition in the bulk. The latter transition takes place in the presence of a metallic surface; in the language of surface critical phenomena this corresponds to the so-called extraordinary transition. The lines of the surface and the extraordinary transitions meet at the special transition point. To elucidate universal properties at different transitions on the phase diagram, we employ renormalization group methods and compute the corresponding surface critical exponents using $\varepsilon$-expansion.
\end{abstract}

\maketitle

\section{Introduction} \label{sec:intro}

Nodal semimetals such as Weyl and Dirac semimetals have garnered substantial attention since their recent discovery owing to their remarkable electronic properties and potential applications in various fields. They represent a class of three-dimensional materials characterized by gapless electronic excitations appearing when linear band crossings occur at isolated points in the Brillouin zone and thus they can be viewed as higher dimensional analogs of the celebrated graphene~\cite{Armitage2018}. In crystals where either the inversion or the time reversal symmetry is broken, bands are generally non-degenerate so that the crossing gives rise to Weyl nodes without fine tuning~\cite{Xu:2015a,Xu:2015b}. If the symmetry under simultaneous inversion and time reversal holds, the bands are two-fold degenerate so that the crossing leads to Dirac nodes~\cite{Liu:2014, Neupane:2014, Borisenko:2014}. The nodal semimetals exhibit peculiar properties arising from their relativistic low-energy excitations. These include novel responses to applied electric and magnetic fields, e.g. due to the chiral anomaly, the appearance of unusual surface states which could be topologically protected~\cite{Yang2011, Burkov2016, Son:2013}, and non-Anderson phase transitions in the presence of disorder.

In a semi-infinite semimetal, scattering from the boundary creates surface modes with energies near the bulk band crossing. While the general properties of these emergent surface states are dictated by the topology and symmetries, their precise form is determined by the microscopic boundary conditions (BCs), which describe how the different wave function components mix upon reflection of the excitation from the boundary~\cite{hashimoto_boundary_2017, faraei_greens_2018, ShtankoLevitov2018, Brillaux:2021}. In Weyl semimetals, topologically protected surface-bound states appear in the form of Fermi arcs that connect the surface projections of Weyl nodes with opposite chirality~\cite{Xu:2015a, Xu:2015b}. In Dirac semimetals, the surface sates may consist of doubled arcs that bridge the surface projections of Dirac nodes~\cite{Xu2015-2}. However in this case, for some BCs, the surface Fermi line can shrink down to a point producing a Dirac cone in the surface spectrum~\cite{VolkovEnaldiev:2016, ShtankoLevitov2018, Brillaux:2021}.

The presence of quenched disorder can strongly modify the behavior of clean materials and lead to Anderson localization~\cite{Abrahams:2010} and continuous Anderson transitions, which include metal-insulator transitions as well as transitions between different topological phases such as the integer quantum Hall plateau transitions~\cite{Evers:2008}. Anderson transitions lack a conventional order parameter: the average local density of states (LDOS) is nonsingular across the transitions. Instead, critical behavior at Anderson transitions is exhibited by transport coefficients and the whole distribution of the LDOS. Different moments of the LDOS have independent scaling behavior reflecting the multifractal nature of critical wave functions described by a continuum of critical exponents, the so-called multifractal spectrum.

A different type of disorder-induced quantum phase transition was identified in nodal semimetals~\cite{Syzranov:2018}, wherein a strong enough disorder drives the semimetal from the clean (ballistic) behavior towards a diffusive metal. This transition is described by the Gross-Neveu model in the replica limit $\mathcal{N} \to 0$~\cite{Roy:2014,Louvet:2016} or its supersymmetric variant~\cite{Syzranov:2015b}. The average LDOS at the nodal point plays the role of an order parameter, since it becomes nonzero above a critical disorder strength~\cite{Sbierski:2014, Sbierski:2016, Fradkin:1986, Roy:2016b, Goswami:2011, Hosur:2012, Ominato:2014, Chen:2015, Altland:2015:2016, SzaboRoy:2020}. This transition has been intensively studied using both numerical simulations~\cite{Kobayashi:2014, Sbierski:2015, Liu:2016, Bera:2016, Fu:2017, Sbierski:2017} and analytical methods~\cite{Roy2016,Balog:2018, Louvet:2017, Sbierski:2019, Klier:2019}. Similar to Anderson transitions, critical wave functions (at zero energy) are multifractal, but the distributions of non-self-averaging quantities, such as the LDOS, are much more narrow compared to those at Anderson transitions. Another difference is that the average LDOS is smooth across many Anderson transitions contrary to the typical LDOS, which vanishes in the localized phase~\cite{Janssen1998}. At the semimetal-diffusive metal transition, both the typical and the average LDOS at the nodal point vanish in the semimetal phase but grow in the metallic phase with different exponents~\cite{Brillaux:2020}. Effects of rare events have been also discussed, and the possibility for an avoided quantum criticality was much debated~\cite{Holder:2017, Gurarie:2017, Nandkishore:2014, Pixley:2016, Pixley:2016c, Wilson:2017, Ziegler:2018, Buchhold:2018, Buchhold:2018-2, PixleyWilson:2021,PiresSantosAmorim2021}.

The disorder-driven quantum phase transitions in a semi-infinite geometry are less well studied. In the theory of conventional continuous phase transitions in semi-infinite spin systems one distinguishes three boundary universality classes: the ordinary, the extraordinary, and the special~\cite{Cardy1996}. For the ordinary transition the bulk and the boundary order simultaneously, while at the extraordinary transition the bulk orders in the presence of already ordered boundary. The special transition corresponds to a multicritical point where the lines of the ordinary, the extraordinary, and the surface transitions meet~\cite{lubensky_critical_1975_a, lubensky_critical_1975, Diehl-The-Theory-1997}. Anderson transitions in the presence of boundaries were studied in Refs.~\cite{Subramaniam:2006, Mildenberger-Boundary-2007, Obuse-Multifractality-2007, Obuse-Corner-2008, Obuse-Boundary-2008, Subramaniam-Boundary-2008, Babkin-Generalized-2023, Babkin-Boundary-2023}, with the focus on multifractality of critical wave functions. Multifractal spectra were found to be modified near boundaries. We note that this modification was studied only at the ordinary boundary critical point. The possibility of extraordinary and special boundary Anderson transitions is an open issue that is interesting to consider and study.

How disorder modifies surface states and affects the bulk criticality in nodal semimetals is much less known. Numerical simulations show that while Fermi arcs in Weyl semimetals are robust against weak bulk disorder, they hybridize with nonperturbative bulk rare states as the strength of disorder gradually increases, and completely dissolve into the emerging metallic bath at the bulk transition~\cite{Slager2017, Wilson:2018}. Perturbative calculations also show that the surface states in generic Dirac materials are protected from surface disorder due to a slow decay of the states from the surface~\cite{ShtankoLevitov2018}.

\begin{figure}[b!t]
\centering
\includegraphics[scale=0.8]{\PathToFigures 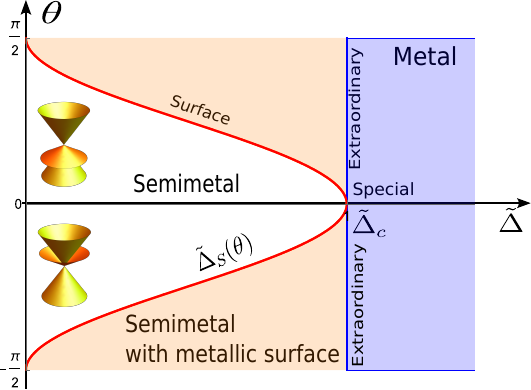}
\caption{Phase diagram in the $(\tilde{\Delta},\theta)$ plane of a semi-infinite Dirac semimetal, where $\tilde{\Delta}$ is the dimensionless strength of disorder and  $\theta$ is the angle parametrizing the time reversal boundary conditions (see Sec.~\ref{sec:model}). The surface transition line and the extraordinary transition line are given by $\tilde{\Delta}_{S} = \cos^2 \theta$ and $\tilde{\Delta}_{c} = 1$, respectively. For $\tilde{\Delta} < \tilde{\Delta}_{S}$ both the surface and the bulk are in the semimetal phase, for $\tilde{\Delta}_{S}< \tilde{\Delta} < \tilde{\Delta}_{c} $ metallic eigenstates populate the surface, and for $\tilde{\Delta}> \tilde{\Delta}_{c} $ the bulk becomes a diffusive metal as well. The lines of surface and extraordinary transitions meet at the multicritical point of the special transition. }
\label{fig:phase_diagram}
\end{figure}

In Ref.~\cite{Brillaux:2021} two of us studied effects of weak disorder on the surface states produced by generic boundary conditions in nodal semimetals using a local version of the self-consistent Born approximation (SCBA)~\cite{Klier:2019}. We investigated the full phase diagram in the presence of a surface and found that for the BCs leading to the Fermi arcs on the surface of Weyl and Dirac semimetals the disorder driven transition belongs to the extraordinary class, i.e. the bulk becomes metallic when the surface is already metallic, with a finite LDOS at the Fermi energy. However, contrary to the bulk criticality where in the replica limit $\mathcal{N} \to 0$ there is no difference between Weyl and Dirac fermions, the surface criticality can be different for Weyl and Dirac fermions. In particular, for Dirac fermions there is a class of time-reversal invariant BCs, which can be parametrized by the angle $\theta$,  where the Fermi surface shrinks to a point on the boundary which hosts single-cone Dirac surface states. It turns into a metallic state at a finite strength of disorder which is lower than the critical strength of disorder in the bulk. This leads to a much richer phase diagram shown in Fig.~\ref{fig:phase_diagram} which exhibits the special transition where the lines of extraordinary transition and surface transition meet.

In the present paper we study universal properties at various transitions in this phase diagram using renormalization group (RG) methods. Recently the surface critical behavior of interacting fermions in the Gross-Neveu universality class has been studied using conformal field theory methods in~\cite{Giombi:2022, Herzog:2023}. However, as we will show, unlike the bulk case, the critical theory describing the behavior of disordered Dirac fermions in the presence of a surface is \emph{not} the Gross-Neveu model in the replica limit $\mathcal{N} \to 0$. Thus the critical exponents obtained in Refs.~\cite{Giombi:2022, Herzog:2023} are different from those found by us below and given in Eqs.~(\ref{eq-eta-rho-s})-(\ref{eq-beta-s-3}).

The paper is organized as follows. In Sec.~\ref{sec:model} we introduce the model of a semi-infinite Dirac semimetal, discuss the BCs satisfying different symmetries and compute the Green's function and the LDOS profile in a clean system. Section~\ref{sec:disorder} summarizes the phase diagram of a disordered semi-infinite Dirac semimetal for time-reversal invariant BCs computed using the SCBA and RG methods. In Sec.~\ref{sec:field-theory-RG-general} we construct the corresponding replicated effective field theory averaged over different disorder distributions and discuss its renormalization and define the surface critical exponents. In Secs.~\ref{sec:bulk}-\ref{sec:surface} we study the bulk, special, extraordinary, and surface transitions. Section~\ref{sec:conclusions} summarizes our results. Some technical details are presented in the Appendixes.

%%%%%%%%%%%%%%%%%%%%%%%%%%%%%%% Phase Diagram %%%%%%%%%%%%%%%%%%%%%%%%

%%%%%%%%%%%%%%%%%%%%%%%%%%%%%%%%%%%%%%%%%%%%%%%%%%%%%%%%%%%%%%%%%%%%

\section{Semi-infinite Dirac semimetal}
\label{sec:model}

\subsection{Hamiltonian and boundary conditions}

The low-energy Hamiltonian which describes noninteracting electrons in a clean three-dimensional (3D) Dirac semimetal can be written as
\begin{align}
\hat{H}_0 = -i\hbar v_F \, \alpha_i \partial_i,
\label{eq:Ham0}
\end{align}
where the $4 \times 4$ Dirac matrices $\alpha_i $ satisfy the anticommutation relations: $\alpha_i \alpha_j + \alpha_j \alpha_i = 2 \delta_{ij}$ [$i, j = 1, 2, 3$ and $(x^1, x^2, x^3) = (x, y, z)$]. In what follows we set the Fermi velocity $\hbar v_F=1$, choose the Weyl representation $\alpha_j = \tau_3 \otimes \sigma_j$ where $\tau_j$ and $\sigma_j$ are Pauli matrices, and use $\tau_0 $, $\sigma_0 $ for the identity matrix or leave it implicit.

In the case of a semi-infinite semimetal filling the half-space $z>0$ the Hamiltonian~\eqref{eq:Ham0} has to be supplemented with a boundary condition (BC) for spinor wave functions $\psi$ at the surface $z=0$. In general case  it can be written as~\cite{ShtankoLevitov2018}
\begin{align}
M \psi|_{z=0} = \psi|_{z=0},
\label{eq:BC-1}
\end{align}
where $M$ is a unitary Hermitian matrix, i.e. $M^2 = 1$, constrained by the condition that Hamiltonian~\eqref{eq:Ham0} is Hermitian in a semi-infinite space $z\ge 0$, i.e. $\langle \psi_1 | \hat{H}_0 \psi_2 \rangle = \langle \hat{H}_0 \psi_1 | \psi_2 \rangle $ for arbitrary $\psi_1$ and $\psi_2$ satisfying the BC~\eqref{eq:BC-1}. This gives that $\alpha_3$ anticommutes with $M$,
\begin{align}
\{ \alpha_3, M \} = \alpha_3 M + M \alpha_3 =0.
\label{eq:BC-current}
\end{align}
The condition~\eqref{eq:BC-current} implies that the $z$ component of the current $J_i = \psi^\dagger \alpha_i \psi$ (normal to the boundary) vanishes at $z=0$. As we demonstrate in Appendix~\ref{app:BC}, the most general matrix satisfying Eq.~\eqref{eq:BC-current} can be parametrized by four angles $\theta \in[-\frac{\pi}2,\frac{\pi}2] $, $\gamma \in[0,\frac{\pi}2]$, and $\phi, \psi \in[-\pi,\pi)$ as follows:
\begin{align}
M_{\mathrm{general}} &= \cos\theta \cos\gamma \, \tau_0 \otimes (\cos\psi \, \sigma_1 + \sin\psi \, \sigma_2)
\nonumber \\
& \quad - \sin\theta \cos\gamma \, \tau_3 \otimes (\sin\psi \, \sigma_1 - \cos\psi \, \sigma_2)
\nonumber \\
& \quad - \sin\theta \sin\gamma \, (\sin\phi \, \tau_1 - \cos\phi \, \tau_2) \otimes \sigma_0
\nonumber \\
& \quad + \cos\theta \sin\gamma \, (\cos\phi \, \tau_1 + \sin\phi \, \tau_2) \otimes \sigma_3.
\label{eq:M-general}
\end{align}
This can be considered as a generalization to three dimensions of the BCs derived for graphene in Refs.~\cite{Akhmerov-Boundary-2008,McCann-Falco:2004}.
Here we limit our consideration to the case of BCs which are rotationally invariant in plane $(x,y)$. As shown in Appendix~\ref{app:BC-symmetry}, such BCs are also conformally invariant, and are given by Eq.~\eqref{eq:M-general} with $\gamma = \frac{\pi}{2}$, which reduces to a two-parameter family (after the shift $\phi \to \phi + \pi$)
\begin{align}
M_\mathrm{3D conf} &= \sin\theta \, (\sin\phi \, \tau_1 - \cos\phi \, \tau_2) \otimes \sigma_0
\nonumber \\
& \quad - \cos\theta \, (\cos\phi \, \tau_1 + \sin\phi \, \tau_2) \otimes \sigma_3.
\label{eq:M-conformal}
\end{align}
The BCs~\eqref{eq:BC-1} with $M_{\mathrm{general}}$ and $\gamma \neq \frac{\pi}{2}$ lead to the emergence of surface states forming the so-called Fermi rays which connect the projection of the Dirac point on the surface with infinity similar to the Fermi arcs in Weyl semimetals where they connect the projection of the two Weyl points on the surface.

The family of conformally invariant BCs with matrices $M$ given by Eq.~\eqref{eq:M-conformal} contains the time-reversal invariant BCs which satisfy $[\mathcal{T}, M] = 0$. They can be obtained by setting $\phi = \pm \pi/2$ and without loss of generality written as
\begin{align}
M_\theta &= \sin\theta \, \tau_1 \otimes \sigma_0 - \cos\theta \, \tau_2 \otimes \sigma_3.
\label{eq:M-T-family}
\end{align}
The matrix~\eqref{eq:M-T-family} can be rotated to~\eqref{eq:M-conformal} by $ M_\mathrm{3D conf} = U_\phi M_\theta U_\phi^{-1} $ with the unitary matrix
\begin{equation}
U_\phi = \left[ \cos\left(\frac{\pi}4-\frac{\phi}2 \right) \tau_0 + i \sin\left(\frac{\pi}4-\frac{\phi}2 \right) \tau_3 \right] \otimes \sigma_0.
\end{equation}
Since this leaves the Hamiltonian unchanged, all the results which we will derive for the matrix~\eqref{eq:M-T-family} can also be applied to BCs with matrix~\eqref{eq:M-conformal}, \textit{e.g.}, the DOS profiles and the critical exponents in both cases are the same. Note that the only conformal invariant BCs which satisfy the charge-conjugation symmetry, $[\mathcal{C}, M] = 0$, are $M_{\theta = 0}$ which, as we will see later, corresponds to the special transition.

\subsection{Dirac surface states}

We now solve the time-independent, massless Dirac equation $\hat{H}_0\psi = \epsilon \psi$ with Hamiltonian~\eqref{eq:Ham0} and BC~\eqref{eq:M-T-family}. We can use the translational invariance along the boundary to perform the Fourier transform in the directions parallel to the surface, $ \vec{r} = (x,y) \to \vec{k} = (k_1,k_2)$, and look for solutions in the form
\begin{align}
\psi(\mathbf{r}) &= \psi_{\vec{k}}(z) e^{i \vec{k}\cdot \vec{r}}.
\end{align}
Introducing $\vec{\alpha}= (\alpha_1,\alpha_2)$, we get
\begin{align}
(-i \, \alpha_3 \partial_3 + \vec{\alpha} \cdot \vec{k})\psi_{\vec{k}}(z)
&= \epsilon_k \psi_{\vec{k}}(z),
\label{eq:Dirac1}
\\
M_\theta \psi_{\vec{k}}(0) &= \psi_{\vec{k}}(0).
\label{BC}
\end{align}
We look for solutions that are localized on the surface $z=0$, and, therefore, substitute the ansatz
\begin{align}
\psi_{\vec{k}}(z) & = \chi_{\vec{k}} e^{-\mu_k z}. \label{eq:Dirac-solution00}
\end{align}
Rotation invariance in the $x$-$y$ plane ensures that the eigenvalues $\epsilon_k$ and the inverse decay lengths of the surface states $\mu_k \geq 0$ depend only on $k = |\vec{k}|$. The Fourier transformed equation and the boundary condition become a system of algebraic equations:
\begin{align}
&(i \mu_k \alpha_3 + k \cos\varphi \, \alpha_1 + k \sin\varphi \, \alpha_2 - \epsilon_k) \chi_{\vec{k}} = 0,
\\
&
(M_\theta - 1) \chi_{\vec{k}} = 0,
\label{BC-theta}
\end{align}
where we denoted by $\varphi$ the angle specifying the direction of the vectors $\vec{k}$ in the $x$-$y$ plane:
\begin{align}
k_1 &= k \cos\varphi, & k_2 &= k \sin\varphi.
\end{align}
Then the eigenstates of the Dirac equation~\eqref{eq:Dirac1} with BC~\eqref{eq:M-T-family} which are localized on the surface read
\begin{align}
\chi_{\vec{k}} = \frac1{\sqrt{2}} \big( 1,\mp e^{i (\theta +\varphi )},-i e^{i \theta },
\mp i e^{i \varphi }
\big)^{\mathrm{T}}.
\label{eq:Dirac-solution1}
\end{align}
Their energy and inverse penetration length are
\begin{align}
\epsilon_k &= \mp k \cos\theta,
\label{eq:esp-k}
\\
\mu_k &= \pm k \sin\theta.
\label{eq:mu-k}
\end{align}
Notice that the solutions are confined to the surface only when $\mu_k>0$, and thus, the eigenstates with the upper sign exist only for $0 <\theta< \frac{\pi}2$ and have only negative energy, while the solutions with the lower sign exist only for $-\frac{\pi}2< \theta < 0$ and have only positive energy. Thus, these surface states form a single cone with the Fermi velocity $v_F\cos\theta$ (smaller than the Fermi velocity in the bulk $v_F$) that extends in either the electron or hole side for negative and positive values of $\theta$, respectively.  These surface cones are schematically depicted in Fig.~\ref{fig:phase_diagram} for negative and positive $\theta$ together with the bulk Dirac cone.

The surface states completely dissolve in the bulk Dirac continuum and disappear for $\theta=0$ that is a 3D generalization of the armchair edge of graphene. On the contrary, for $\theta=\pm \frac{\pi}2 $, the surface states form non-dispersing flat bands with $\epsilon_k = 0$ but a finite penetration length $\mu^{-1}_k=k^{-1}$. This can be viewed as a generalization of a flat band of spin-polarized states localized at zigzag edges of graphene~\cite{Fujita-Wakabayashi-Nakada-Kusakabe-JPhys-Japan:1996}. In Appendix~\ref{appendix:slab} we show that these surface states survive in a slab geometry of thickness $L$ once $k L \sin\theta \gg 1$.

\subsection{Green's function}

The boundary breaks translational invariance along the perpendicular direction. Consequently, the retarded Green's function $G_0(\bm{x},\bm{x}',\epsilon+i0^+)$ depends not only on the distance $\bm{x}-\bm{x}'$ between the points but also on the distances of the points to the surface, i.e. $z$ and $z'$. Performing the Fourier transform in the directions parallel to the surface and keeping the real space coordinates in $z$ direction we define the Green's function as
\begin{align}
\label{eq:Gdef}
(\hat{H}_0 - \epsilon)G_0(\vec{k},z,z',\epsilon) = \delta(z-z'),
\end{align}
and impose the BC
\begin{align}
\label{eq:BC-2}
M_\theta G_0(\vec{k},0,z,\epsilon)=G_0(\vec{k},0,z,\epsilon).
\end{align}
Notice that in Eq.~\eqref{eq:Gdef} we change the sign of $\mathcal{G}$ with respect to the standard notation to match our definition of the Green's function in the field theory. The Green's function can be split into bulk and surface parts,
\begin{align}
\label{eq:G-full}
G_0 = G_b + G_s,
\end{align}
where
\begin{widetext}
\begin{center}
\vspace{-7mm}
\begin{align}
\label{eq:G-b-1}
G_b(\vec{k},z,z',\epsilon) &= \Big(\frac{\vec{\alpha} \cdot \vec{k} + \epsilon}{2q_{k,\epsilon}}
+ \frac{i}{2} \mathrm{sign}(z-z') \alpha_3 \Big) e^{-q_{k,\epsilon} |z-z'|},
\\
\label{eq:G-s-1}
G_s(\vec{k},z,z',\epsilon) &= -\frac{ e^{-q_{k,\epsilon} (z+z')}}{2 (q_{k,\epsilon} \cot\theta +\epsilon)}
\bigg[\frac{k^2}{q_{k,\epsilon}} + \Big(\frac{\epsilon}{q_{k,\epsilon}} + i \alpha_3 + \frac{\tau_2\otimes \sigma_3}{\sin\theta} \Big) (\vec{\alpha} \cdot \vec{k} )
+ \frac{q_{k,\epsilon} \tau_1 \otimes \sigma_0 + \epsilon \tau_2 \otimes \sigma_3 }{\sin\theta} \bigg],
\end{align}
\end{center}
\end{widetext}
and we introduced $q_{k,\epsilon} = \sqrt{k^2-\epsilon^2}$. For $k < |\epsilon|$ we need to distinguish the retarded and advanced GFs by shifting the energy $\epsilon$ off the real axis. If we denote by $s=\pm 1$ the sign in the shift of the energy: $\epsilon \to \epsilon + i s 0^+$, then the appropriate analytic continuation of the GFs~\eqref{eq:G-b-1} and~\eqref{eq:G-s-1} is achieved by the replacement
\begin{align}
q_{k,\epsilon} &\to -is \epsilon t,
&
t &\equiv \sqrt{1 - k^2/\epsilon^2} \in [0,1].
\end{align}

The surface part of the Green's function~\eqref{eq:G-s-1} can be expressed in terms of the bulk part~\eqref{eq:G-b-1} in two limiting cases: (i) in the limit of zero energy $\epsilon=0$; (ii) in the limit of $\theta=0$ when there are no surface states. In both cases the Green's function reduces to
\begin{align}
G_0(\vec{k},z,z',\epsilon) = G_b(\vec{k},z,z',\epsilon)+ U_\theta G_b(\vec{k},-z,z',\epsilon),
\label{eq-Greens-U0}
\end{align}
where we introduce
\begin{align}
U_\theta = -\frac1{\cos\theta}\ \tau_2 \otimes \sigma_3 -i \tan \theta \ \tau_3 \otimes \sigma_3,
\end{align}
such that $U_{\theta=0} = M_{\theta=0}$. Note that the Green's function at zero energy $\epsilon=0$ can be expressed in the real space as
\begin{align}
G_0(\bm{x},\bm{x}') = \frac{i}{S_3} \left[ \frac{\alpha_j (x-x')_j}{|\bm{x}-\bm{x}'|^d}+ U_\theta \frac{\alpha_j (\bar{x}-x')_j}{|\bar{\bm{x}}-\bm{x}'|^d} \right],
\label{eq:G-real-space}
\end{align}
where we defined $\bm{x} = (\vec{r}, z)$, $\bar{\bm x} = (\vec{r},-z)$, and $S_d = 2\pi^{d/2}/[\Gamma(d/2)]$ is the area of the unit sphere in $d$ dimensions. The Green's function similar to Eq.~\eqref{eq:G-real-space} was derived in~\cite{McAvity:1993} for the BC of the form~\eqref{eq:BC-1} with a Hermitian $M$ such that $M^2 = 1$, and $M$ commutes with $\alpha_1$,$\alpha_2$ and anticommutes with $\alpha_3$.

\subsection{Profile of the density of states}

%%%%%%%%%%%%%%%%%%% Figure 1 LDOS %%%%%%%%%%%%%%%%%%%%%%%%%%%%%%%%%%%%%%%
\begin{figure*}[htpb]
\includegraphics[scale=0.9]{\PathToFigures 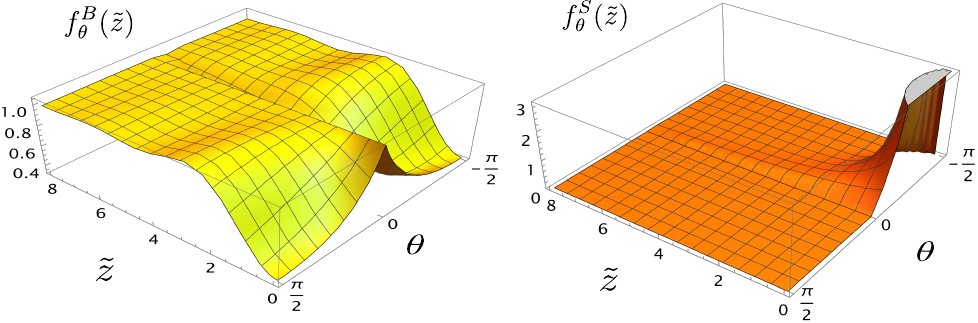}
\caption{Dimensionless LDOS as a function of $\tilde{z}= \frac{\epsilon z}{\hbar v_F} $ and $\theta$ for $\epsilon>0$. (\textbf{Left panel}) $f_\theta^B(\tilde{z})$ is the contribution from the extended states which exhibit the Friedel oscillations decaying for $\tilde{z} \to \infty$ so that $f_\theta^B(\infty)=1$. (\textbf{Right panel}) $f_\theta^S(\tilde{z})$ is the contribution from the surface states localized at $z=0$ whose energy is positive for $\theta<0$ and negative for $\theta>0$.
}
\label{fig:rho-BS}
\end{figure*}
%%%%%%%%%%%%%%%%%%%%%%%%%%%%%%%%%%%%%%%%%%%%%%%%%%%%%%%%%%%%%%%%%%%%%%%%%%

We are now in the position to compute the LDOS profile using the standard relation between the LDOS and the retarded Green's function which reads
\begin{align} \label{eq:LDOS-def}
\rho_{\theta}(z,\epsilon) = \frac1{\pi} \mathrm{Im} \!\! \int \frac{d^2k}{(2\pi)^2} \lim\limits_{z'\to z} \mathrm{Tr} \, G_0(\vec{k},z,z',\epsilon+i0^+).
\end{align}
Substituting the Green's function~\eqref{eq:G-full} we obtain
\begin{align}
\rho_{\theta}(z,\epsilon) &= \frac2{\pi} \mathrm{Im}\!\! \int \!\! \frac{d^2k}{(2\pi)^2}
\bigg[ \frac{\epsilon}{q_{k,\epsilon}}
-\frac{k^2 e^{-2 q_{k,\epsilon} z}}{q_{k,\epsilon}(q_{k,\epsilon} \cot\theta + \epsilon )}\bigg]_{\epsilon \to \epsilon+i0^+}.
\label{eq:LDOS-1}
\end{align}
We can split the LDOS into the bulk and surface parts resulting from the states delocalized and localized at the surface, respectively
\begin{align}
\rho_{\theta}(z,\epsilon) &= \rho_{\theta}^B(z,\epsilon) + \rho_{\theta}^S(z,\epsilon).
\label{eq:LDOS-2}
\end{align}
The first term under the integral on the r.h.s of~\eqref{eq:LDOS-1} gives the $z$-independent LDOS in the bulk far from the surface
\begin{align}
\rho_{\theta}^B(z\to \infty,\epsilon) &= \frac2{\pi} \int_0^{|\epsilon|} \frac{k d k}{2\pi} \frac{|\epsilon|}{\sqrt{\epsilon^2-k^2}} = \frac{\epsilon^2}{\pi^2}.
\label{eq:LDOS-b}
\end{align}
The second term under the integral on the r.h.s of~\eqref{eq:LDOS-1} can be split into two parts. The integration over $0<k<|\epsilon|$ gives the correction to the bulk LDOS resulting from the distortions of the bulk states near the surface, i.e. the so-called Friedel oscillations. Then the full bulk LDOS can be written as $\rho_{\theta}^B(z,\epsilon) = \frac{\epsilon^2}{\pi^2} f_{\theta}^B(\epsilon z)$ with
\begin{align}
f_{\theta}^B(\tilde z)= 1+ \mathrm{Im}\, \int_0^1 dt
\frac{ \left(1-t^2\right) e^{2 i t \tilde{z} }}{t \cot \theta + i}.
\label{eq:LDOS-F}
\end{align}

The integration over $k>|\epsilon|$ yields the surface states contribution to the LDOS and can be evaluated using the Sokhotski-Plemelj theorem. The pole on the integration contour exists only if $\epsilon \tan\theta < 0$ and the corresponding residue yields $\rho_{\theta}^S(z,\epsilon) = \frac{\epsilon^2}{\pi^2} f_{\theta}^S(\epsilon z)$ with
\begin{align}
f_{\theta}^S(\tilde z) = \frac{\pi|\tan(\theta )| }{\cos^2 \theta}
[1 - \Theta_\mathrm{H} (\tilde{z} \tan \theta )] e^{2 \tilde{z} \tan\theta},
\label{eq:LDOS-surface}
\end{align}
where $\Theta_\mathrm{H}(x)$ is the Heaviside step function. Restoring the physical units we obtain
\begin{align}
{\rho}_{\theta}^A(z,\epsilon) &= \frac{\epsilon^2}{(\pi \hbar v_F)^2}
f_{\theta}^A \left(\frac{\epsilon z}{\hbar v_F}\right),
&
A &= B,S.
\end{align}
The typical value of the Fermi velocity in Dirac semimetals is $\hbar v_F \simeq 1 \div 10 \, \unit{\electronvolt\angstrom}$. The bulk and surface contributions to the LDOS profile for different $\theta$ and $\epsilon>0$ are shown in Fig.~\ref{fig:rho-BS}. The time reversal symmetry implies that $f_{\theta}^A(\tilde{z}) = f_{-\theta}^A(-\tilde{z})$.

\section{Effect of disorder in self-consistent Born approximation}
\label{sec:disorder}

The presence of disorder creates a random time-independent potential $V(\bm{x})$, which is a scalar in the simplest case, so that the low energy Hamiltonian becomes
\begin{align}
\hat{H} = -i \, \alpha_i \partial_i + V(\bm{x}).
\label{eq:Ham1}
\end{align}
We assume that the distribution of disorder potential is translationally invariant, isotropic, and
Gaussian with the mean value and variance given by
\begin{align}
\overline{V(\bm{x})} &= 0,
&
\overline{V(\bm{x}) V(\bm{x'})} &= \Delta \delta (\bm{x}-\bm{x}'),
\end{align}
where the overbar indicates the disorder average.

The self-consistent Born approximation (SCBA) is the simplest approximation which captures the effects of disorder on the Dirac semimetal, at least qualitatively. We now briefly summarize the results of SCBA obtained by two of us for the disordered Dirac semimetal in Ref.~\cite{Brillaux:2021}. The resulting phase diagram in the plane ($\Delta$, $\theta$) is shown in Fig.~\ref{fig:phase_diagram}.

Far from the surface, the semimetal bulk undergoes a phase transition towards a diffusive metal at $\tilde{\Delta}_c=1$, where $\tilde{\Delta} = \Delta \Lambda/(4\pi)$ is the dimensionless disorder strength and $\Lambda$ is a UV cutoff. In the diffusive metal phase the LDOS at zero energy grows with disorder strength $\tilde{\Delta}$ as ${\rho}(z\to\infty,\epsilon=0)=(\tilde{\Delta}-\tilde{\Delta}_c)^\beta$, where the exponent $\beta=1$ is independent of $\theta$. In the semimetal phase the bulk LDOS at zero energy vanishes; however, depending on $\theta$ it can be finite on the surface for $\tilde{\Delta}_S(\theta) < \tilde{\Delta} \le \tilde{\Delta}_c$. In the later case the LDOS decays away from surface exponentially, $\rho(z,\epsilon=0)\sim e^{-z/\xi}$, for $\tilde{\Delta}_S(\theta) <\tilde{\Delta} < \tilde{\Delta}_c$ and algebraically, $\rho(z,\epsilon=0)\sim (z\Lambda)^{-1}$, at criticality $\tilde{\Delta} = \tilde{\Delta}_c$. Here the correlation length $\xi\sim |\tilde{\Delta} - \tilde{\Delta}_c|^{-\nu}$ can be associated with the mean free path and the SCBA predicts $\nu=1$. For weaker disorder $\tilde{\Delta} < \tilde{\Delta}_S(\theta)$, the LDOS at zero energy vanishes identically in the whole system, and thus, both bulk and surface remain in the semimetal phase.

Using the local SCBA it was shown in Ref.~\cite{Brillaux:2021} that $\tilde{\Delta}_S=(\cos \theta)^2$. Indeed, at $\theta=0$ there are no surface states in the clean sample (see the previous section) and as a result the surface becomes metallic simultaneously with the bulk. At $\theta=\pm\frac{\pi}2$ there is a flat band at zero energy on the surface so that it has non-zero LDOS even in the clean limit. Thus, the line of the surface transition, if exists, has to pass through these points.

In terms of surface critical phenomena the semimetal exhibits the so-called extraordinary transition at $\tilde{\Delta}=1$ for $\theta \in [-\frac{\pi}2,0)\cup(0,\frac{\pi}2] $. The line of the extraordinary transition meets the line of the surface transition at the multicritical point $(\tilde{\Delta}=1, \theta = 0)$ which corresponds to the special transition. Note that the surface transition can only exist on the boundary of a disordered three-dimensional semimetal, where the surface states have a small overlap with impurities in the bulk, and the electrons can bypass them in three dimensions. This is not the case for strictly two-dimensional disordered Dirac materials where there is no such freedom, and even weak disorder becomes relevant~\cite{Ludwig:1994, Ostrovsky:2006, Fedorenko:2012}. Below we will recover this phase diagram and study the corresponding critical properties using RG methods.

\section{Effective field theory and RG equations}
\label{sec:field-theory-RG-general}

\subsection{Effective action}

Disordered Dirac fermions in a semi-infinite space are described by the Hamiltonian~\eqref{eq:Ham1} with the BCs~\eqref{eq:BC-1}. This boundary condition may also be incorporated into the Hamiltonian of the problem by adding a boundary term of the form $\tilde{M} \delta(z)$. Then the Schr\"{o}dinger equation
\begin{align}
[- i \bm{\alpha} \cdot \nabla + V(\bm{x}) + \tilde{M} \delta(z)] \psi = i \partial_t \psi
\end{align}
can be integrated across the boundary, and assuming $\psi = 0$ for $z < 0$ this gives
\begin{align}
[-i \alpha_3 + \tilde{M}]\psi|_{z=0} = 0.
\end{align}
Thus including the surface term is equivalent to the explicitly imposed boundary condition~\eqref{eq:BC-1} if we identify
\begin{align}
\tilde{M} = i \alpha_3 M.
\end{align}
Let us construct the corresponding Euclidean action. Since we are studying noninteracting fermions, the action can be written at a fixed Matsubara frequency $\omega$ as~\cite{hashimoto_boundary_2017}
\begin{align}
S &= \int_{z>0} d^3 x \, \bar{\psi}(\bm{x})(-i \, \alpha_i \partial_i - i \omega + V(\bm{x}) ) \psi(\bm{x}) \nonumber \\
& \quad + \int d^{2} r \bar{\psi}(\vec{r}) \tilde{M} \psi(\vec{r}),
\label{eq:action-1}
\end{align}
where $\bar{\psi}$ and $\psi$ are spinor Grassmann fields. The last term in the action~\eqref{eq:action-1} implies that the corresponding Green's functions satisfy the boundary condition~\eqref{eq:BC-1}. In particular, for the BCs with matrix $M$ given by Eq.~\eqref{eq:M-T-family} in the clean limit of $V=0$ we recover the one particle Green's function~\eqref{eq:G-full}-\eqref{eq:G-s-1} with $\epsilon= i\omega$.

In what follows we will use a dimensional regularization of integrals, so we need to generalize the action~\eqref{eq:action-1} to an arbitrary dimension $d$. We introduce coordinates $\bm{x} = (\vec{r},z)$, $z\ge 0$ in a semi-infinite $d$-dimensional space, where $\vec{r}$ lies in the $(d-1)$-dimensional surface $z=0$. We keep the orbital Pauli matrices $\tau_j$, $j = 1, 2, 3$, but generalize the spin Pauli matrices $\sigma$ to a $d$-dimensional Clifford algebra $\sigma_\mu$, $\mu=1,...,d$ and call $\sigma_d=\sigma_z$. This generates $\alpha_\mu =\tau_3 \otimes \sigma_\mu $, $\mu=1,...,d$.

We now apply the replica trick and introduce $\mathcal{N}$ copies of the system denoted by indices $a,b = 1, \ldots, \mathcal{N}$, which are summed over when repeated. The effective replicated action averaged over disorder reads
\begin{align}
\label{eq:action-2}
S &= \int_{z>0} d^d x \, \bar{\psi}_a(\bm{x})(-i \, \alpha_\mu \partial_\mu - i \omega ) \psi_{a}(\bm{x})
\nonumber \\
& \quad + \int d^{d-1} r \bar{\psi}_{a}(\vec{r}) \tilde{M} \psi_a(\vec{r})
\nonumber \\
& \quad - \frac{\Delta}{2} \int_{z>0} d^d x \, \bar{\psi}_a(\bm{x})\psi_{a}(\bm{x}) \bar{\psi}_b(\bm{x})\psi_{b}(\bm{x}).
\end{align}
The observables averaged over disorder can be computed using the action~\eqref{eq:action-2} in the limit of $\mathcal{N} \to 0$.

\subsection{Renormalization and RG equations}

The correlation functions computed perturbatively in the disorder strength $\Delta$ using the action~\eqref{eq:action-2} are UV divergent in $d=2$, which is the lower critical dimension of the model: while any disorder is relevant for $d\le 2$, weak disorder is irrelevant for $d>2$. We will employ dimensional regularization and compute correlation functions in $d=2+\varepsilon$ dimensions, thereby converting UV divergences into poles in $\varepsilon$. To renormalize the theory, we will use the minimal subtraction scheme within the context of dimensional regularization and collect all poles in $\varepsilon$ in the $Z$ factors $Z_{\psi}$, $Z_{\psi s}$, $Z_\omega$, and $Z_\Delta$, so that the renormalized action reads
\begin{widetext}
\vspace{-7mm}
\begin{align}
\label{eq:action-r}
S_R &= \int \!\! \frac{d^{d-1} k}{(2\pi)^{d-1}} \int_0^\infty \!\! dz \, \bar{\psi}_{\alpha}(-\vec{k},z)
\big[ Z_\psi (\vec{\alpha} \cdot \vec {k} - i \alpha_z \partial_z)
- Z_\omega i \omega \big] \psi_{a}(\vec{k},z)
+ \int \!\! \frac{d^{d-1} k}{(2\pi)^{d-1}} \bar{\psi}_{s\,a}(-\vec{k}) \tilde{M} \psi_{s\,a}(\vec{k})
\nonumber \\ &
- \frac{ \mu^{-\varepsilon }\Delta}{K_d} \int \prod_{i=1}^{3} \frac{d^{d-1} k_i}{(2\pi)^{d-1}}
\int_0^\infty dz \, \bar{\psi}_{a}(\vec{k}_1,z)
\psi_{a}(\vec{k}_2,z) \bar{\psi}_{b}(\vec{k}_3,z) \psi_{b}(-\vec{k}_1-\vec{k}_2-\vec{k}_3,z).
\end{align}
Here we have introduced the renormalized bulk and surface fields $\psi(\vec{k},z)$ and $\psi_s(\vec{k},z)$ and the notation $K_d = S_d/(2\pi)^{d}$. The engineering (momentum) dimensions of the fields $\psi(\bm{x})$ and $\psi_s(\bm{x})$ in real space are equal, as can be seen from the first line in Eq.~\eqref{eq:action-2}:
\begin{align}
d_\psi^0 = \frac{1}{2}(d - 1).
\label{engineering-dimensions}
\end{align}
The Fourier transforms in the $d-1$ directions along the surface [between $\psi(\bm{x})$ and $\psi(\vec{k},z)$, as well as between $\psi_s(\bm{r})$ and $\psi_s(\vec{k})$] change the engineering dimensions of the fields and introduce delta functions that reflect the conservation of momentum in correlators of the fields.

The renormalized fields and the dimensionless coupling constant $\Delta$ at the mass scale $\mu$ are related to their bare values denoted by circles as
\vspace{-2mm}
\begin{align}
\mathring{\psi} &= Z_{\psi}^{1/2} \psi,
&
\mathring{\psi_s} &= Z_{\psi_s}^{1/2} {\psi_s},
%\label{eq:Z-factors} \\
&
\mathring{\omega} &= Z_\omega Z_{\psi}^{-1}\omega,
&
\mathring{\Delta} &= \frac{2 \mu^{-\varepsilon }}{K_d} \frac{Z_\Delta}{Z_{\psi}^{2}} \Delta. \label{eq:Z-factors2}
\end{align}
The renormalized correlation function with $n$ bulk fields $\psi$ or $\bar{\psi}$ and $m$ surface fields $\psi_s$ or $\bar{\psi}_s$ is given by
\begin{align}
\mathring{G}^{(n,m)}(\vec{k}_i,z_j; \mathring{\omega},\mathring{\Delta})
&= Z_{\psi}^{n/2} Z_{\psi_s}^{m/2} G^{(n,m)}(\vec{k}_i,z_j; \omega, \Delta, \mu),
\end{align}
where $\vec{k}_i$ stands for $n+m$ momenta of fields $\psi$, $\bar{\psi}$, $\psi_s$, $\bar{\psi}_s$ such that $\sum_i \vec{k}_i=0$ and $z_j$ stands for $n$ $z$-coordinates of the bulk fields $\psi$, $\bar{\psi}$. The $z$ coordinates of $m$ fields $\psi_s$, $\bar{\psi}_s$ are all $0$.
Using the fact that $\mathring{G}^{(n,m)}$ does not depend on the mass scale $\mu$ we derive the RG equations
\begin{align}
\biggl[\mu\frac{\partial}{\partial \mu} - \beta(\Delta)
\frac{\partial}{\partial \Delta} + \frac{n}2 \eta_\psi(\Delta) + \frac{m}2 \eta_{\psi_s}(\Delta)
- \gamma (\Delta) \omega \frac{\partial}{\partial \omega}\biggr]
G^{(n,m)}(\vec{k}_i,z_j; \omega, \Delta, \mu)=0,
\label{eq-RG-1}
\end{align}
where we have introduced the scaling functions
\begin{align}
\beta(\Delta) &= -\mu\frac{\partial \Delta}{\partial \mu} \bigg|_{\mathring{\Delta}},
&
\eta_i(\Delta) &= -\beta(\Delta)\frac{\partial \ln Z_i}{\partial \Delta},
\quad (i=\psi,\psi_s,\omega),
&
\gamma(\Delta) &= \eta_\omega(\Delta)- \eta_\psi(\Delta).
\label{eq:scal-fun-gamma}
\end{align}
Dimensional analysis implies that for arbitrary rescaling factor $\lambda > 0$
\begin{align}
& G^{(n,m)}(\vec{k}_i,z_j; \omega, \Delta, \mu)
= \lambda^{(n + m - 2) d_\psi^0}
G^{(n,m)}\Big(\lambda \vec{k}_i, \frac{z_j}{\lambda};
\lambda \omega, \Delta, \lambda \mu \Big).
\label{G-rescaling}
\end{align}
The factor $\lambda^{-2d_\psi^0} = \lambda^{-(d-1)}$ comes from the delta function expressing the conservation of momentum. Equation~\eqref{G-rescaling} can be rewritten in the infinitesimal form as
\begin{align}
\bigg[\mu\frac{\partial}{\partial \mu} + \sum\limits_i \vec{k}_i \frac{\partial}{\partial \vec{k}_i} -
\sum\limits_j z_j \frac{\partial}{\partial z_j} + \omega \frac{\partial}{\partial \omega}
+ (n + m - 2) d_\psi^0 \bigg] G^{(n,m)}(\vec{k}_i,z_j; \omega, \Delta, \mu) = 0.
\label{eq-inf}
\end{align}
Subtracting Eq.~\eqref{eq-RG-1} from Eq.~\eqref{eq-inf} we arrive at
\begin{align}
\bigg[\beta(\Delta)
\frac{\partial}{\partial \Delta} + \sum\limits_i \vec{k}_i \frac{\partial}{\partial \vec{k}_i} -
\sum\limits_j z_j \frac{\partial}{\partial z_j}
+ [1 + \gamma(\Delta)] \omega \frac{\partial}{\partial \omega}
+ n \Big(d_\psi^0 - \frac{1}{2} \eta_\psi(\Delta) \Big)
+ m \Big(d_\psi^0 - \frac{1}{2} \eta_s(\Delta)\Big) - 2d_\psi^0 \bigg] G^{(n,m)} = 0.
\label{eq-RG-final}
\end{align}
\vspace{-5mm}
%\end{widetext}

The linear partial differential equation~\eqref{eq-RG-final} can be solved using the method of characteristics in which the general solution is written as
\begin{align}
G^{(n,m)}(\vec{k}_i,z_j; \omega, \Delta)
&= A_{nm}(\xi) f_{nm}[\vec{k}_i(\xi), z_j(\xi); \omega(\xi), \Delta(\xi)],
\end{align}
with an arbitrary function $f_{nm}$. The characteristics are the lines in the space of $k_i$, $z_j$, $\omega$, and $\Delta$, along which the solution of~\eqref{eq-RG-final} propagates. These lines can be parametrized by an auxiliary parameter $\xi$ and are determined by the system of ordinary differential equations
\begin{align}
\frac{d \Delta (\xi)}{d \ln \xi} &= \beta[\Delta(\xi)],
&
\frac{d \vec{k}_i (\xi)}{d \ln \xi} &= \vec{k}_i(\xi),
&
\frac{d z_j (\xi)}{d \ln \xi} &= -z_j(\xi),
&
\frac{d \omega (\xi)}{d \ln \xi} &= \big( 1 + \gamma[\Delta(\xi)] \big) \omega(\xi),
\end{align}
with initial conditions $\vec{k}_i(1)=\vec{k}_i$, $z_j(1)=z_j,$ $\omega (1)=\omega$, and $\Delta (1)=\Delta$. The solution along the characteristics propagates according to
\begin{align}
\frac{d \ln A_{nm}(\xi)}{d \ln \xi} &= 2(n + m - 1)d_\psi^0
- n \Big(d_\psi^0 + \frac{1}{2} \eta_\psi[\Delta(\xi)] \Big)
- m \Big(d_\psi^0 + \frac{1}{2} \eta_s[\Delta(\xi)]\Big),
\end{align}
with initial condition $A_{nm}(1)=1$.

We assume that the $\beta$ function has an unstable fixed point (FP) such that
\begin{align}
\beta(\Delta^*) &= 0, & \beta'(\Delta^*) &> 0.
\label{eq:fixpoint0}
\end{align}
In the vicinity of the FP the solutions of the RG equation~\eqref{eq-RG-final} can be written as
\begin{align}
G^{(n,m)}(\vec{k}_i, z_j, \omega) &=
\xi^{2(n + m - 1)d_\psi^0 - n d_\psi - m d_{\psi_s}}
f_{nm}\Big(\vec{k}_i \xi , \frac{z_j}{\xi}; \omega \xi^\mathfrak{z}, \delta \xi ^{1/\nu}\Big),
\label{eq-scal-an-1}
\end{align}
\end{widetext}
where we have introduced the critical exponents $\nu$, $\mathfrak{z}$, $d_\psi$, $d_{\psi_s}$ and $\delta=\Delta-\Delta^*$.
The scaling ansatz ~\eqref{eq-scal-an-1} implies
\begin{align}
\xi &\sim \delta^{-\nu}, & \frac1{\nu} &= \beta'(\Delta^*),
\end{align}
so that we can identify $\xi$ with the correlation length. The dynamical critical exponent is
\begin{align}
\mathfrak{z} = 1 + \gamma(\Delta^*) = \eta_\omega(\Delta^*)- \eta_\psi(\Delta^*),
\label{dynamical-exponent}
\end{align}
and the (renormalized) scaling dimensions of the fields $\psi$ and $\psi_s$ are
\begin{align}
d_i = d_\psi^0 + \frac{1}{2} \eta_{i}(\Delta^*), \quad i=\psi, \psi_s.
\end{align}
Thus, for example, two-point functions at the transition behave as follows: in the bulk
\begin{align}
G^{(2,0)}(r) &\sim \frac1{r^{d-1+\eta}},
&
\eta &= \eta_{\psi}(\Delta^*),
\end{align}
on the surface
\begin{align}
G^{(0,2)}(r) &\sim \frac1{r^{d-1+\eta_{\parallel}}},
&
\eta_{\parallel}=\eta_{\psi_s}(\Delta^*),
\end{align}
and in the direction $z$ (perpendicular to the surface)
\begin{align}
G^{(1,1)}(z) &\sim \frac1{z^{d-1+\eta_{\perp}}},
&
\eta_{\perp}=\frac12(\eta+\eta_\parallel).
\end{align}

\subsection{Renormalization of composite operators}

To calculate the LDOS we have to consider the renormalization of the composite operators
$O(x):=\bar{\psi}(x){\psi}(x)$ in the bulk and $O_s(r):=\bar{\psi}_s(r){\psi}_s(r)$ on the surface.
We define the $Z$ factors which make their insertions in the renormalized Green's function finite,
\begin{align}
\mathring{O} &= Z_\omega Z_\psi^{-1} O,
&
\mathring{O}_s &= Z_{O_s} Z_{\psi_s}^{-1} O_s.
\end{align}
The functions with $l$ insertions of the composite operator $O$ and $p$ insertions of operator $O_s$ renormalize as
\begin{widetext}
\vspace{-3mm}
\begin{align}
\mathring{G}^{(n,m,l,p)}(\vec{k}_i,z_j; \mathring{\omega},\mathring{\Delta})
= Z_{\psi}^{n/2-l} Z_{\psi_s}^{m/2-p} Z_\omega^{l} Z_{O_s}^{p}
G^{(n,m,l,p)}(\vec{k}_i,z_j; \omega, \Delta, \mu).
\end{align}
By analogy with Eq.~\eqref{eq-RG-final} it is easy to obtain an RG equation for $G^{(n,m,l,p)}$. Then we specialize to the LDOS in the bulk by setting $n=m=p=0$, $l=1$, and integrating over $\vec{k}$, which gives
\begin{align}
&\bigg[ \beta(\Delta)
\frac{\partial}{\partial \Delta} - z \frac{\partial}{\partial z} + [1+\gamma (\Delta)] \omega \frac{\partial}{\partial \omega}
+ \eta_\omega(\Delta) - \eta_\psi(\Delta) - 2 d_\psi^0 \bigg] \rho(z,\omega,\Delta) = 0.
\label{eq:DOS-bulk}
\end{align}
The solution of Eq.~\eqref{eq:DOS-bulk} in the vicinity of the FP~\eqref{eq:fixpoint0} gives the LDOS at the distance $z$ from the surface
\begin{align}
\rho(z, \omega, \delta) = \xi^{\mathfrak{z}-d} f\Big(\frac{z}{\xi}, \omega \xi^\mathfrak{z}, \delta \xi^{1/\nu}\Big),
\label{eq:LDOS-profile-1}
\end{align}
where we have used the relation~\eqref{dynamical-exponent}. The LDOS at the Fermi energy far in the bulk ($z\to \infty$) is given by
\begin{align}
\rho( \delta) &\sim \delta^{\beta},
&
\beta &= \nu(d-\mathfrak{z}).
\end{align}
The RG equation for the LDOS on the surface is obtained by choosing $n=m=l=0$, $p=1$, and $z=0$, which gives
\begin{align}
\bigg[\beta(\Delta)
\frac{\partial}{\partial \Delta} + [1+\gamma (\Delta)] \omega \frac{\partial}{\partial \omega}
+ \eta_{\rho_s}(\Delta) - \eta_{\psi_s}(\Delta) - 2 d_\psi^0 \bigg] \rho_s(\omega,\Delta) = 0, \label{eq:DOS-surface}
\end{align}
\end{widetext}
where we defined the scaling function
\begin{align}
\eta_{\rho_s}(\Delta) &= -\beta(\Delta)\frac{\partial \ln Z_{O_s}}{\partial \Delta}.
\label{eq:rho-s-1}
\end{align}
The solution of Eq.~\eqref{eq:DOS-surface} in the vicinity of the FP~\eqref{eq:fixpoint0} gives the surface LDOS
\begin{align}
\rho_s(\omega, \delta) &= \xi^{- 2 d_\psi^0 + \eta_{\rho_s} - \eta_{\psi_s}} f_s( \omega \xi^\mathfrak{z}, \delta \xi^{1/\nu}),
\end{align}
where
\begin{align}
\eta_{\rho_s} &= \eta_{\rho_s}(\Delta^*),
&
\eta_{\psi_s} &= \eta_{\psi_s}(\Delta^*).
\end{align}
The surface LDOS at the Fermi energy is
\begin{align}
\rho_s( \delta) &\sim \delta^{\beta_s},
&
\beta_s &= \nu(d-1-\eta_{\rho_s}+ \eta_{\psi_s}),
\end{align}
where we defined a new surface exponent $\beta_s$.

\section{Renormalization in the bulk} \label{sec:bulk}

The bulk critical exponents have been computed to two-loop order in Refs.~\cite{Syzranov:2015b,Roy:2014,Louvet:2016} using different RG schemes. Here we repeat these calculations within our RG scheme to get the renormalization constants which we will need to study the surface properties.

To renormalize the theory far from the surface we can take the limit $z,z' \to \infty$ keeping $z-z'$ finite. This allows us to perform the Fourier transform also with respect to $z$, and switch from the correlation functions $G$ to irreducible vertex functions $\Gamma$~\cite{Zinn-Justin:1986}. In order to renormalize the theory it is enough to consider only the one-particle vertex function ${\Gamma}^{(2)}(\bm{k})$ at a finite external momentum $\bm{k}$ and the two particle vertex function ${\Gamma}^{(4)}(\bm{k}_i=0)$ at zero external momenta. The one- and two-loop diagrams contributing to these vertex functions and the corresponding integrals computed using dimensional regularization within minimal subtraction scheme are shown in Appendix~\ref{Appendix-bulk}.  For the one-particle bare vertex function we obtain
\begin{align}
& \mathring{\Gamma}^{(2)}(\bm{k}, \mathring{\omega}) = \alpha_\mu k_\mu \bigg[1 + \frac{K_d^2}4 \mathring{\omega}^{2\varepsilon} \mathring{\Delta}^2 \frac1{\varepsilon} \bigg]
\nonumber \\
& - i \mathring{\omega} \bigg[1
- \frac{K_d}2 \mathring{\omega}^{\varepsilon} \mathring{\Delta} \frac{2}{\varepsilon} + \frac{K_d^2}4 \mathring{\omega}^{2\varepsilon} \mathring{\Delta}^2 \left(\frac{6}{\epsilon ^2}+\frac{4}{\epsilon }\right) \bigg],
\end{align}
while the two-particle bare vertex function reads
\begin{align}
\mathring{\Gamma}^{(4)}(\bm{k}_i=0,\mathring{\omega}) &= \mathring{\Delta} - \frac{K_d}2 \mathring{\omega}^{\varepsilon} \mathring{\Delta}^2  \left(\frac4{\varepsilon} + 6 \right) \nonumber \\
& + \frac{K_d^2}4 \mathring{\omega}^{2\varepsilon} \mathring{\Delta}^3\left(\frac{16 }{\varepsilon ^2}+\frac{54}{\varepsilon }\right).
\end{align}
Using the minimal subtraction scheme we require
\begin{align}
& Z_{\psi} \mathring{\Gamma}^{(2)}(\bm{k}, \mathring{\omega},\mathring{\Delta}) = \mathrm{finite}, \\
& Z^2_{\psi} \mathring{\Gamma}^{(4)}(\bm{k}_i=0, \mathring{\omega},\mathring{\Delta}) = \mathrm{finite},
\end{align}
where the bare parameters $\mathring{\omega}$ and $\mathring{\Delta}$ are expressed in terms of the renormalized ones as in Eq.~\eqref{eq:Z-factors2}.
To two-loop order we obtain
\begin{align}
& Z_{\psi} = 1 - \frac{\Delta^2}{\varepsilon}, \\
& Z_\omega = 1 + \frac{2 \Delta }{\varepsilon} + \frac{6 \Delta^2 }{\varepsilon^2}, \\
& Z_{\Delta} = 1 + \frac{4 \Delta }{\varepsilon} + \Delta^2\left(\frac{16}{\varepsilon^2} + \frac{2 }{\varepsilon}\right)
\end{align}
The corresponding bulk scaling functions can be computed using Eqs.~\eqref{eq:scal-fun-gamma} which give
\begin{align}
\beta (\Delta) &= - \varepsilon \Delta + 4 \Delta^2 + 8 \Delta^3 + O(\Delta^4), \label{eq:beta-fun-2loop} \\
\eta_\psi (\Delta) &= - 2\Delta^2 + O(\Delta^3),
\\
\eta_\omega (\Delta) &= 2\Delta + O(\Delta^3),
\\
\gamma (\Delta) &= 2\Delta + 2\Delta^2 + O(\Delta^3).
\end{align}
Solving the FP equation~\eqref{eq:fixpoint0} with the $\beta$ function~\eqref{eq:beta-fun-2loop} we get
\begin{align}
\Delta^*=\frac{\varepsilon }{4} -\frac{\varepsilon ^2}{8} + O(\varepsilon^3).
\label{eq:fixpoint-2loop}
\end{align}
This gives the critical exponents to two-loop order as
\begin{align}
\frac1{\nu} &= \varepsilon + \frac{\varepsilon^2}{2} + O(\varepsilon^3),
\\
\mathfrak{z} &= 1+ \frac{\varepsilon}{2} -\frac{\varepsilon^2}{8} + O(\varepsilon^3),
\\
\beta &= \nu(d - \mathfrak{z}) = \nu \Big(1+ \frac{\varepsilon}{2} + O(\varepsilon^2) \Big),
\\
\eta &= -\frac{\varepsilon^2}{8} + O(\varepsilon^3),
\\
d_\psi &= \frac12 (d-1+\eta_\psi) = \frac12 + \frac{\varepsilon}{2} -\frac{\varepsilon^2}{16} + O(\varepsilon^3),
\end{align}
which coincide with those found in Refs.~\cite{Syzranov:2015b, Roy:2014,Louvet:2016}.

\section{Special transition} \label{sec:special}

In this section we focus on the special transition taking place at $\theta=0$, when both the bulk and the surface become metallic at the same disorder strength $\Delta_c$ whose dimensionless value is given to two-loop order by Eq.~\eqref{eq:fixpoint-2loop}.

To determine the renormalization factor $Z_{\psi_s}$ one can consider either the surface correlation function $\mathring{G}^{(0,2)}(\vec{k},z_1=0,z_2=0)$ or the mixed bulk-surface correlation function $\mathring{G}^{(1,1)}(\vec{k},z_1=0,z_2=z)$. Here and afterward we explicitly keep the coordinate $z=0$ of the surface fields for transparency. To check that the same $Z$ factors make all correlation functions finite simultaneously we renormalize both these functions. The surface correlation function can be expressed graphically to one loop order as
\begin{align*}
\includegraphics[scale=0.52]{\PathToFigures 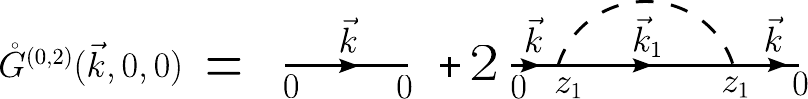},
\end{align*}
where the arrow lines stand for the bare Green's function~\eqref{eq:G-full} and the dashed line for the disorder vertex $\mathring{\Delta}$. Integrating out the internal coordinate $z_1$ we arrive at
\begin{widetext}
\begin{center}
\begin{align}
\mathring{G}^{(0,2)}(\vec{k},0,0) &= (1 + M_0) \frac{\vec{\alpha}\cdot \vec{k}+i\mathring{\omega}}{2q}
+ \frac{\mathring{\Delta}}{2q} (1+M_0) \bigg[ \frac12 \vec{\alpha}\cdot \vec{k} \Big (1-\frac{\mathring{\omega}^2}{q^2}\Big)
+ i\mathring{\omega} \Big(1-\frac{\mathring{\omega}^2}{2q^2}\Big) \bigg] \int_{\vec{k}_1}\frac1{q_1}, \label{eq:G20-1}
\end{align}
\end{center}
where $q=\sqrt{k^2+\mathring{\omega}^2}$ and $q_1=\sqrt{k_1^2+\mathring{\omega}^2}$. The bulk-surface correlation function to one-loop order can be expressed as
\begin{align*}
\includegraphics[scale=0.52]{\PathToFigures 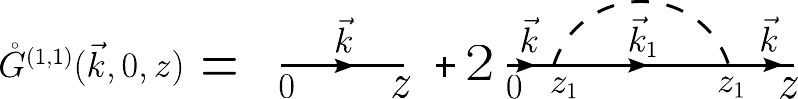},
\end{align*}
which gives
\begin{align}
\mathring{G}^{(1,1)}(\vec{k},0,z) &= (1+M_0)\bigg[\frac{\vec{\alpha}\cdot \vec{k} + i\mathring{\omega}}{2q}-\frac{i \alpha_z}{2}\bigg] e^{-q z}
\nonumber \\ &
+ \frac{\mathring{\Delta}}{8q} (1+M_0) \bigg[\vec{\alpha}\cdot \vec{k} \Big( 1- 2(1+ q z)\frac{\mathring{\omega}^2 }{q^2} \Big)
+  i\mathring{\omega} \Big(3-2 (1+q z)\frac{\mathring{\omega}^2 }{q^2} \Big)
- i q\alpha_z \Big( 1-\frac{2 \mathring{\omega}^2 z}{q} \Big) \bigg] e^{-q z} \int_{\vec{k}_1}\frac1{q_1}.
\label{eq:G11-1}
\end{align}
The one-loop integral over internal momentum $\vec{k}_1$ in Eqs.~\eqref{eq:G20-1} and \eqref{eq:G11-1} can be computed using dimensional regularization as
\begin{align}
\int_{\vec{k}_1}\frac1{q_1}=\int \frac{d^{d-1}k_1}{(2\pi)^{d-1}}\frac1{\sqrt{k_1^2+\omega^2}} = - K_{d-1}\frac{\omega^\varepsilon}{\varepsilon}.
\end{align}
We renormalize the correlation functions by imposing the conditions
\begin{align}
Z_{\psi_s}^{-1} \mathring{G}^{(0,2)}(\vec{k},0,0; \mathring{\omega}, \mathring{\Delta}) &= \mathrm{finite},
&Z_{\psi}^{-1/2} Z_{\psi_s}^{-1/2} \mathring{G}^{(1,1)}(\vec{k},0,z; \mathring{\omega}, \mathring{\Delta})
&= \mathrm{finite},
\end{align}
where the bare parameters $\mathring{\omega}$ and $\mathring{\Delta}$ are expressed in terms of the renormalized ones~\eqref{eq:Z-factors2}. Both conditions yield the same result:
\begin{align}
 Z_{\psi_s} = 1- \frac{2\Delta}{\varepsilon} + O(\Delta^2). \label{eq-psi-s}
\end{align}
Thus, to one loop order we get
\begin{align}
\eta_{\psi_s} = \eta_{\parallel}=2\eta_{\perp} = - 2\Delta^* + O(\Delta^{*2})= - \frac{\varepsilon}{2}+O(\varepsilon^2).
\label{eq-eta-psi-s}
\end{align}
The fact that both correlation functions~\eqref{eq:G20-1} and~\eqref{eq:G11-1} can be made finite simultaneously at arbitrary arguments $\vec{k}$ and $\omega$ with the same $Z$ factors is nontrivial and proves consistency of our calculations.

To find the factor $Z_{O_s}$ for the surface composite operator we can consider the surface correlation function $\mathring{G}^{(0,2,0,1)}(k,z_1=0,z_2=0,0)$ with an insertion of $O_s$ at zero momentum $\vec{k}_{O_s}=0$, which to one-loop order reads
\begin{align*}
\includegraphics[scale=0.5]{\PathToFigures 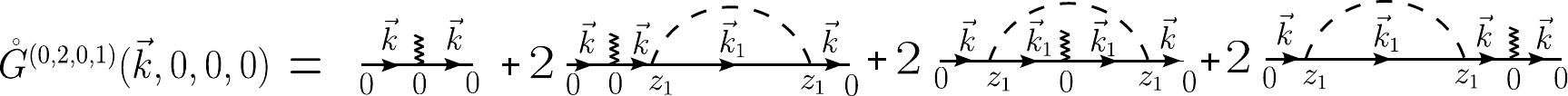}.
\end{align*}
Here wavy lines denote the insertions of the composite operator $O_s$. Evaluating the Feynman diagrams we arrive at
\begin{align}
\mathring{G}^{(0,2,0,1)}(k,0,0,0) &= \frac1{q^2} (1+M_0) \bigg[ {i\mathring{\omega}(\vec{\alpha}\cdot \vec{k}) -\mathring{\omega}^2} \bigg]
+ \frac{\mathring{\Delta}}{q^2} (1+M_0) \bigg[ i \mathring{\omega} ( \vec{\alpha}\cdot \vec{k})
\Big(2-\frac{\mathring{\omega}^2}{q^2}\Big)
- \mathring{\omega}^2 \Big(\frac52-\frac{\mathring{\omega}^2}{q^2}\Big) \bigg] \int_{\vec{k}_1} \frac1{q_1},
\end{align}
Similarly, the bulk-surface correlation function $\mathring{G}^{(1,1,0,1)}(k,z_1=0,z_2=z,0)$ with an insertion of an $O_s$ composite operator can be expressed graphically as
\begin{align*}
\includegraphics[scale=0.5]{\PathToFigures 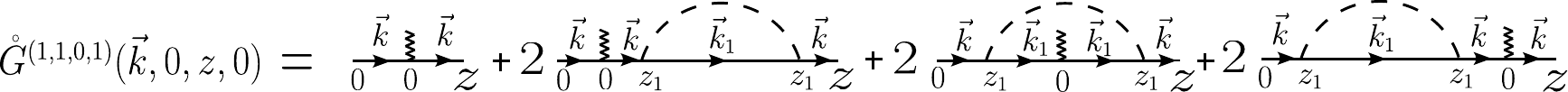},
\end{align*}
which gives
\begin{align}
\mathring{G}^{(1,1,0,1)}(\vec{k},0,z,0) &= (1+M_0)\bigg[ (\vec{\alpha}\cdot \vec{k}) \Big( \frac{i \mathring{\omega}} {q^2} -\frac{i}{4q} \alpha_z
-\frac{1}{4q} \tau_x \otimes \sigma_0 \Big) + \frac{q^2-2\mathring{\omega}^2}{q} + \mathring{\omega}\alpha_z
\bigg] e^{-q z}
\nonumber \\ &
+ \frac{\mathring{\Delta}}{8q}\, (1+M_0) \bigg[ (\vec{\alpha} \cdot \vec{k}) \bigg(2 i \frac{\mathring{\omega}}{q}
\Big(9-2 (2+ q z) \frac{\mathring{\omega}^2}{q^2} \Big)
- \Big(5-2 (1+q z) \frac{\mathring{\omega}^2}{q^2} \Big) \tau_x \otimes \sigma_0 \bigg)
\nonumber \\ &
+ q\Big(5 - 2(9+q z) \frac{\mathring{\omega}^2}{q^2} +4 (2+q z) \frac{\mathring{\omega}^4}{q^4}\Big)
+ \mathring{\omega} \Big(7-2 \left(1+q z\right) \frac{\mathring{\omega}^2}{q^2}\Big)\alpha_z \bigg]
e^{-q z} \int_{\vec{k}_1} \frac1{q_1}.
\end{align}
\end{widetext}
To renormalize these correlation functions with insertions we require
\begin{align}
Z_{O_s}^{-1} \mathring{G}^{(0,2,0,1)}(\vec{k},0,0,0; \mathring{\omega}, \mathring{\Delta}) &= \mathrm{finite},
\nonumber \\
Z_{\psi}^{-1/2} Z_{\psi_s}^{1/2} Z_{O_s}^{-1}
\mathring{G}^{(1,1,01)}(\vec{k},0,z,0; \mathring{\omega}, \mathring{\Delta}) &= \mathrm{finite},
\end{align}
where the bare parameters $\mathring{\omega}$ and $\mathring{\Delta}$ are expressed in terms of the renormalized ones~\eqref{eq:Z-factors2}. To lowest order in $\Delta$, renormalization of both correlation functions gives
\begin{align}
 Z_{O_s} = 1- \frac{6\Delta}{\varepsilon}+ O(\Delta^2). \label{eq-O-s}
\end{align}
From Eqs.~~\eqref{eq:rho-s-1} and \eqref{eq-O-s} we find
\begin{align}
 \eta_{\rho_s} = -6\Delta^* = -\frac32 \varepsilon +O(\varepsilon^2). \label{eq-eta-rho-s}
\end{align}
The surface dynamic exponent is
\begin{align}
 \mathfrak{z}_{s} = 1+ \eta_{\rho_s} - \eta_{\psi_s} = 1-\varepsilon +O(\varepsilon^2). \label{eq-z-surf-1}
\end{align}
Thus, to one loop order the surface states in three dimensions ($\varepsilon = 1$) have very weak dispersion $\omega \sim k^{\mathfrak{z}_s}$ with $\mathfrak{z}_{s}\approx 0$. The critical behavior of the LDOS on the boundary is described by the surface critical exponent
\begin{align}
\beta_{s} = \nu(d-\mathfrak{z}_{s}) = \nu \big[1+2\varepsilon + O(\varepsilon^2) \big]. \label{eq-beta-s-2}
\end{align}
The ratio of the surface and bulk exponent independent of $\nu$ reads
\begin{align}
\frac{\beta_{s}}{\beta } \approx \frac{ 1+2 \varepsilon }{1+\frac12 \varepsilon}
= 1 + \frac{3}{2} \varepsilon + O(\varepsilon^2),
\label{eq-beta-s-3}
\end{align}
which can be helpful since numerical estimates of the exponent $\nu$ usually have larger error bars than those for the exponent $\beta$~\cite{Louvet:2016}.

Let us make some remarks on the relation between disordered Dirac semimetals and the Gross-Neveu model. It has been shown~\cite{Roy:2014,Louvet:2016,Syzranov:2015b} that the bulk critical behavior of disordered Dirac and Weyl semimetals can be derived from the bulk critical behavior of the Euclidean Gross-Neveu model
\begin{align}
S &= - \int \! d^{d} x \Big[\sum_{a=1}^\mathcal{N} \bar{\psi}_a(\bm{x})(\gamma^\mu_{\mathrm{E}} \partial_\mu )  \psi_{a}(\bm{x})
\nonumber \\ & \quad
+ \frac{g}{2} \sum_{a,b=1}^{\mathcal{N}} \bar{\psi}_a(\bm{x})\psi_{a}(\bm{x}) \bar{\psi}_b(\bm{x})\psi_{b}(\bm{x}) \Big].
\label{eq:action-GN}
\end{align}
in the limit of the number of fermion flavors $\mathcal{N}=0$. Indeed, if we change variable $\bar{\psi} \to - i\bar{\psi} $ in~\eqref{eq:action-2} and omit for the moment the surface and mass terms we obtain
\begin{align}
S &= - \int \! d^{d} x \Big[\sum_{a=1}^\mathcal{N} \bar{\psi}_a(\bm{x})(\alpha_\mu \partial_\mu )  \psi_{a}(\bm{x})
\nonumber \\ & \quad
- \frac{\Delta}{2} \sum_{a,b=1}^{\mathcal{N}} \bar{\psi}_a(\bm{x})\psi_{a}(\bm{x}) \bar{\psi}_b(\bm{x})\psi_{b}(\bm{x}) \Big].
\label{eq:action-disDirac2}
\end{align}
Since the Euclidean matrices $\gamma^\mu_{\mathrm{E}}$ satisfy the same anticommutation relations as $\alpha_\mu$ both models~\eqref{eq:action-GN} and~\eqref{eq:action-disDirac2} lead to the same bulk critical behavior. Note that $d$ in Eq.~\eqref{eq:action-GN} is not the space dimension as in Eq.~\eqref{eq:action-disDirac2} but the space-time dimension, and the positive sign of $\Delta$ in Eq.~\eqref{eq:action-disDirac2} corresponds to attraction, while the positive sign of $g$ in Eq.~\eqref{eq:action-GN} corresponds to repulsive interactions. The  Gross-Neveu model has been studied in the high energy physics up to four loop order~\cite{Wetzel:1985, Ludwig:1987, Gracey:2016}, so  the bulk critical behavior of the disordered Dirac fermions can be obtained from previously known results to order~$\varepsilon^4$.

However, the imposition of the BCs at $z=0$ as $M \psi =\psi$ breaks the correspondence between the two models. In semi-infinite systems $z>0$, the same matrix $M$ can give different results for the disorder Dirac fermions and the Gross-Neveu model since $\alpha_\mu$ and $\gamma_E^\mu$ are different and $M$ introduces an explicit dependence of the results on their form. Indeed, the boundary condition with $M_{\theta=0}$ was recently used in Refs.~\cite{Giombi:2022, Herzog:2023} to study the boundary critical behavior of conformal field theories of interacting fermions in the Gross-Neveu universality class. The surface critical exponents derived in Refs.~\cite{Giombi:2022, Herzog:2023} in the limit of $\mathcal{N}=0$ are different from those we obtained here.

\section{Extraordinary transition} \label{sec:extraordinary}

Unlike the special and the ordinary transitions, the RG description of the extraordinary transition encounters serious obstacles even for spin systems~\cite{Diehl-book:1986}. Its current understanding still relies on the mean-field approximation and exact results for spherical spin models~\cite{BrayMoore:1977}.

The case of a Dirac semimetal in a half space is even more complicated than that of a spin system. In the latter, the ordered surface and the extraordinary transition (the bulk transition in the presence of an ordered boundary) exist at the mean-field level. On the other hand, the semimetal phase with a metallic surface is non-perturbative in the disorder strength, and so the extraordinary transition cannot be studied in the usual framework of perturbative RG. Indeed, repeating the calculations which we have done in the previous section for $\theta \ne 0$ we find that the model is not renormalizable: new divergences are generated by the RG flow. They cannot be absorbed into the existing $Z$ factors and can be attributed to the presence of a nonperturbative nonzero surface LDOS at the extraordinary transition. Nevertheless, we can extract some predictions performing calculations in the real space.

The one-loop Green's function on the surface ($z_1=z_2=0$) at zero energy can be written in real space for arbitrary $\theta$ as
\begin{align}
 \includegraphics[scale=0.5]{\PathToFigures 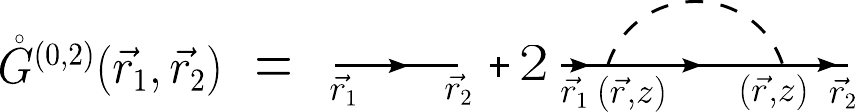}
\end{align}
where the first term is the bare Green's function given by Eq.~\eqref{eq:G-real-space},
\begin{align}
\delta^{(0)} \mathring{G}^{(0,2)}(\vec{r}_{12}) = \frac{i}{S_d} (1 + U_{\theta})
\frac{\vec{\alpha} \cdot \vec{r}_{12}}{r_{12}^d},
\label{eq-G-0}
\end{align}
where $\vec{r}_{12} = \vec{r}_{1}-\vec{r}_{2}$ and $r_{12} = |\vec{r}_{12}|$. The one loop correction can be split into the two integrals:
\begin{widetext}
\vspace{-8mm}
\begin{center}
\begin{align}
\delta^{(1)} \mathring{G}^{(0,2)}(\vec{r}_{12}) = \frac{i \mathring{\Delta}}{S_d^3}\Big(\frac{2}{\cos^2\theta}-1 \Big)
\int d^{d-1}r \int\limits_0^{\infty} dz
\frac{(2z)^{2-d} (1 + U_\theta) \vec{\alpha} \cdot \vec{r}_{12}}
{[(z^2+(\vec{r}-\vec{r}_1)^2)(z^2+(\vec{r}-\vec{r}_2)^2)]^{d/2}},
\label{eq-dG-1}
\end{align}
\end{center}
and
\begin{align}
\delta^{(2)} \mathring{G}^{(0,2)}(\vec{r}_{12}) = - \frac{ 2 \mathring{\Delta}}{S_d^3} \frac{\sin\theta}{\cos^3\theta}
\int d^{d-1}r \int\limits_0^{\infty} dz \, (2z)^{1-d}
\frac{ \{(1+U_\theta) [\vec{\alpha} \cdot (\vec{r}-\vec{r}_1)][ \vec{\alpha} \cdot (\vec{r}-\vec{r}_2)] +2 z^2 \} (1 + M_\theta)}
{[(z^2+(\vec{r}-\vec{r}_1)^2)(z^2+(\vec{r}-\vec{r}_2)^2)]^{d/2}}. \label{eq-dG-2}
\end{align}
The integral in Eq.~\eqref{eq-dG-1} is calculated in Appendix~\ref{Appendix-integrals}. Expanding the numerical prefactor in small $\varepsilon =d-2$ to lowest order we get
\begin{align}
\delta^{(1)} \mathring{G}^{(0,2)} (\vec{r}_{12}) &=
\frac{i \mathring{\Delta} }{4 \pi S_d} \Big(\frac{2}{\cos^2\theta}-1 \Big)
\frac{(1+U_\theta) \vec{\alpha} \cdot \vec{r}_{12}}{r_{12}^{2d-2}}
\Big(-\frac{2}{ \varepsilon} + \gamma_{\mathrm{E}} +\log (\pi ) + O(\varepsilon) \Big).
\label{eq-dG-1-2}
\end{align}
Using that to lowest order in $\varepsilon$ the bare disorder is $\mathring{\Delta}= 2\Delta/K_d$ we can combine the correction~\eqref{eq-dG-1-2} with~\eqref{eq-G-0}, expanding it around the leading asymptotics $\vec{\alpha} \cdot \vec{r}_{12}/r_{12}^d$ as
\begin{align}
\mathring{G}^{(0,2)}(\vec{r}_{12}) &= \frac{i}{S_d} (1+U_\theta) \frac{\vec{\alpha} \cdot \vec{r}_{12}}{r_{12}^d} \bigg[1 + \Delta \Big(\frac{2}{\cos^2\theta}-1 \Big)
\Big( -\frac{2}{\varepsilon }+ \log \left(r_{12}^2 \right) +  2 \gamma_{\mathrm{E}} -2 \log 2 \Big) \bigg].
\label{eq-G-1-3}
\end{align}
The $\log$-term in Eq.~~\eqref{eq-G-1-3} gives us the anomalous dimension of the surface scaling behavior. Exponentiating it in Eq.~\eqref{eq-G-1-3} we get
\begin{align}
G^{(0,2)}(\vec{r}_1,\vec{r}_2) = \frac{i}{S_d} (1+U_\theta) \frac{\vec{\alpha} \cdot \vec{r}_{12}}{r_{12}^{d+\eta_{\parallel}}} \label{eq:delta1G}
\end{align}
where, to one loop order,
\begin{align}
\eta_{\parallel} &= - 2\Big(\frac{2}{\cos^2\theta}-1 \Big) \Delta^* + O(\Delta^{*2})
= - \Big(\frac{2}{\cos^2\theta}-1 \Big) \frac{\varepsilon}{2} + O(\varepsilon^2). \label{eq:eta-extraordinary}
\end{align}
For $\theta =0$ this exponent coincides with that at the special transition and grows with $\theta$ diverging at $\theta=\frac{\pi}2$ corresponding to the surface flat band.

In contrast with extraordinary transitions in spin systems where the critical exponents do not depend on the bare BC, here the critical exponent $\eta_{\parallel}$ depends continuously on the BC parameter $\theta$. As we show in Appendix~\ref{app:BC-symmetry}, the BC~\eqref{eq:BC-1} is conformally invariant for any parameter $\theta$, while in spin systems a general mixed (Robin) BC involves a surface mass $m_s$ that describes the modification of the surface coupling strength and breaks conformal invariance. As a result, the mass $m_s$ flows under RG to its fixed-point values $m_s = \pm \infty$ at the ordinary and extraordinary transitions~\cite{Diehl-Why-2020}, and the surface critical exponents at the extraordinary transition do not depend on the bare values of $m_s$. Instead, in the disordered Dirac systems we find a line of fixed points corresponding to different values of parameter $\theta$, which does not renormalize.

Let us now discuss the second contribution given by Eq.~\eqref{eq-dG-2}. As shown in Appendix~\ref{Appendix-integrals} it can be simplified to
\begin{align}
\delta^{(2)} \mathring{G}^{(0,2)}(\vec{r}_{12}) &= - \frac{ 4 \mathring{\Delta}}{S_d^3} \frac{\sin\theta}{\cos^3\theta} \frac{\Gamma(d)}{\Gamma^2(d/2)}
\int d^{d-1}r \int\limits_0^{\infty} dz
\int\limits_0^1 d \beta \,
\frac{(2z)^{1-d} (1+M_\theta) [r^2+z^2 - \beta(1-\beta) r_{12}^2 ]  }
{[r^2+z^2 + \beta(1-\beta) r_{12}^2]^{d} [\beta(1-\beta)]^{1-d/2} }.
\label{eq:delta2G}
\end{align}
\end{widetext}
For $\vec{r}_{12} \neq 0$, the integral~\eqref{eq:delta2G} converges for $\frac32 < d < 3$ and evaluates to
\begin{align}
\delta^{(2)} \mathring{G}^{(0,2)} (\vec{r}_{12})
&= -\frac{ \mathring{\Delta} }{ S_d } \frac{\sin\theta}{\cos^3\theta} \,
\frac{ \Gamma \big(\tfrac{3-d}{2}\big) \Gamma \big(d-\tfrac{3}{2}\big)}{2 \pi ^{d/2} } \frac{1+M_\theta }{r_{12}^{2d-3}}.
\label{eq:delta2G-2}
\end{align}
Naively, the correction~\eqref{eq:delta2G-2} is subdominant with respect to the leading scaling behavior~\eqref{eq:delta1G} for $d>2$. However, this term has a different structure from the bare two-point function, that is, renormalization of the model requires adding counterterms of the form absent in the bare model. Therefore, the model is not renormalizable for $\theta \neq 0$ in $d=2$.

If the correction~\eqref{eq:delta2G} to the Green's function had an imaginary part, it would give rise to a nonzero surface LDOS at the extraordinary transition. The integral~\eqref{eq:delta2G} diverges at small $z$ for $d>\frac32$ for $\vec{r}_{12}=0$. Introducing a UV cutoff $z>2\pi/\Lambda $ we find that its trace is given by
\begin{align}
\mathrm{Tr} \delta^{(2)} \mathring{G}^{(0,2)} (0) &=
- \frac{ \mathring{\Delta} \Lambda^{1+2\varepsilon} }{S_d^2} \frac{\sin\theta}{\cos^3\theta} \frac{2^{7-3 d} \pi ^{3-2 d} \Gamma \left(\frac{d-1}{2}\right)^2}{(2 d-3) \Gamma (d-1)}.
\end{align}
The result is real, and thus, the surface LDOS vanishes at zero energy in perturbation theory in the disorder strength $\mathring{\Delta}$. A non-zero surface LDOS is generated only at a finite disorder strength $\Delta_s$ at the surface transition, and the semimetal phase with metallic surface (see Fig.~\ref{fig:phase_diagram}) is intrinsically non-perturbative. Obtaining a non-zero surface LDOS requires either the SCBA discussed in Sec.~\ref{sec:disorder} or the RG approach to the surface states developed in the next section.

Nevertheless the profile of the LDOS away from the surface at the extraordinary transition can be found from the solution of Eq.~\eqref{eq:LDOS-profile-1} and reads
\begin{align}
\rho(z) &= \frac1{z^{d-\mathfrak{z}}},
\end{align}
where $\mathfrak{z}$ is the bulk dynamical exponent.

We emphasize that the above calculations do not provide a complete description of the extraordinary transition in disordered Dirac fermions since they do not take into account nonperturbative effects for disorder strengths larger than the surface critical disorder $\Delta > \Delta_s$.

\section{Surface transition} \label{sec:surface}

The critical exponents for the surface transition in spin systems coincide with the critical exponents of the corresponding $(d - 1)$-dimensional system. Indeed, once we integrate out the bulk degrees of freedom, the bare Green's function in the bulk $1/(\tau+k ^2)$, where $\tau$ is the bulk reduced temperature and $\vec{k} \in \mathbb{R}^{d}$, becomes $1/(\tau_s+k^2)$, where $\tau_s$ is the surface reduced temperature and $\vec{k} \in \mathbb{R}^{d-1}$~\cite{BrayMoore:1977}. This is due to the fact that the bulk remains massive when the surface is critical, and thus, the bulk can be effectively decoupled from it. The situation is completely different in the case of a Dirac semimetal, where the bulk remains massless at the surface transition.

Using Eq.~\eqref{eq-Greens-U0} or Eq.~\eqref{eq:G-real-space}, the two-point Green's function on the surface at zero energy can be expressed as
\begin{align}
G_0(\vec{k}) = (1+U_{\theta}) \frac{\vec{\alpha}\cdot \vec{k}}{2k}.
\label{eq:G-surf}
\end{align}
We now construct an effective field theory of the surface transitions. Since the matrix~\eqref{eq:G-surf} is not invertible, it is convenient to switch from the four-component spinor field $\psi(\vec{r})$ to the two-component spinor field
$\Psi(\vec{r})=\{\Psi_1(\vec{r}),\Psi_2(\vec{r})\}$ such that $\psi(\vec{r}) = X \Psi(\vec{r})$ with
\begin{align}
X = \left(
\begin{array}{cc}
 \frac{i}{\sqrt{2}} & 0 \\
 0 & \frac{i}{\sqrt{2}} \\
 \frac{e^{i \theta }}{\sqrt{2}} & 0 \\
 0 & -\frac{e^{-i \theta }}{\sqrt{2}} \\
\end{array}
\right).
\end{align}
This matrix satisfies $X^{\dagger} X = \sigma_0 $ and $(M_\theta-1)X=0$, i.e. the field $\Psi(\vec{r})$ satisfies the BCs~\eqref{eq:BC-1} by construction. We can now write the effective theory of the surface transition as
\begin{align}
S_{\mathrm{surf}} &= \sum\limits_{a=1}^{\mathcal{N}} \int d^{d-1} k \bar{\Psi}_a(-\vec{k}) \mathcal{G}^{-1}(\vec{k}) \Psi_a(\vec{k})
\nonumber \\ &
- \frac{\Delta_s}2 \sum\limits_{a,b=1}^{\mathcal{N}} \int d^{d-1}r \bar{\Psi}_a(\vec{r}) \Psi_a(\vec{r}) \bar{\Psi}_b(\vec{r})\Psi_b(\vec{r}),
\label{eq:H-surf}
\end{align}
where we have introduced the surface interaction term with an effective strength $\Delta_s$ due to scattering of surface states by impurities close to the surface. The Green's function $\mathcal{G}(\vec{k}) = X^{\dagger} G_0(\vec{k}) X $ has the following form
\begin{align}
\mathcal{G}(\vec{k}) = (1- i \sigma_z \tan \theta ) \frac{\vec{\sigma}\cdot \vec{k}}{k}. \label{eq:G-surf-eff}
\end{align}
The dimensional analysis of the theory~\eqref{eq:H-surf} shows that the critical dimension is $d=1$ and weak disorder is irrelevant for $d>1$. To investigate effects of strong disorder we apply the Wilsonian RG method (shell integration in momentum space). To one loop order this gives
\begin{align}
-m \partial_m \Delta_s &= - (d-1) \Delta_s + \frac{2}{\cos^2\theta} \Delta_s^2.
\end{align}
This RG flow has a non-Gaussian fixed point
\begin{align}
\Delta_s^* &= \frac12 (d-1) {\cos^2\theta},
\end{align}
that is unstable and corresponds to the surface transition. In the plane ($\theta$, $\Delta$) shown for $d=3$ in Fig.~\ref{fig:phase_diagram} the line of these fixed points describing the surface transition connects the special transition point and the point ($\theta=\pi/2$, $\Delta=0$) corresponding to the surface flat band which becomes metallic for arbitrary weak disorder similar to the strictly 2D Dirac fermions, e.g. in the graphene. Linearizing around the surface fixed point we calculate the in-plane correlation length exponent at the surface transition for $0<\theta<\pi/2$ as
\begin{align}
\nu_s = \frac1{d-1},
\end{align}
which becomes $\nu=1/(d-2)$ at the special transition for $\theta=0$ while at $\theta=\pi/2$ one can expect an exponential behavior.

\section{Conclusions} \label{sec:conclusions}

In this paper, we have studied a non-Anderson, dis\-or\-der-driven quantum phase transition in a semi-infinite Dirac semimetal with a flat boundary. We considered the most general boundary conditions (BCs) and then narrowed them down to the ones that respect time-reversal invariance and conformal symmetry and can be parametrized by a single angle $\theta$. In clean systems, general BCs give rise to Fermi arcs on the surface, while the conformally invariant BCs lead to rotationally invariant surface states with linear dispersion and a point-like Fermi surface at zero energy.

Studying the large scale behavior of the semi-infinite disordered Dirac fermions as a function of the angle $\theta$ and the disorder strength, we obtained a rich phase diagram: the surface of the system becomes metallic at a critical disorder that is weaker than that for the semimetal-diffusive metal transition in the bulk. The latter transition then takes place in the presence of a metallic surface. In the language of surface critical phenomena this corresponds to the so-called extraordinary transition. The lines of the surface and the extraordinary transitions meet at the special transition point.

To elucidate the universal properties across the phase diagram, we applied renormalization group (RG) methods and computed the corresponding critical exponents. We showed, that unlike the bulk case, the surface critical properties of disordered Dirac fermions are not the same as the surface critical properties of the Gross-Neveu model in the zero replica limit. While the special and the surface transitions can be described perturbatively in disorder from the semimetal phase, the extraordinary transition cannot be studied within the usual framework of perturbative RG since the semimetal phase with a metallic surface is intrinsically nonperturbative in the disorder strength. A complete description of the extraordinary transition in disordered Dirac fermions remains a challenging task.

It would be also interesting to study the non-perturbative effects on the surface criticality and in particular on the surface transition~\cite{Holder:2017, Gurarie:2017, Nandkishore:2014, Pixley:2016, Pixley:2016c, Wilson:2017, Ziegler:2018, Buchhold:2018, Buchhold:2018-2, PixleyWilson:2021, PiresSantosAmorim2021}. We hope that these studies can be generalized to the more complicated case of nodal semimetals with Fermi arcs on the surface and even to the disordered semi-infinite nodal loop semimetals~\cite{BurkovHookBalents:2011, SilvaEtal:2023, ZhuSyzranov:2023}.

Furthermore, it is interesting to reconsider Anderson transitions in semi-infinite disordered electronic systems from the perspective of surface critical phenomena.

%%%%%%%%%%%%%%%%%%%%%%%%%%%%%%%%%%%%%%%
\emph{\label{sec:acknow} Acknowledgments.} --
%%%%%%%%%%%%%%%%%%%%%%%%%%%%%%%%%%%%%
We would like to thank David Carpentier, Lucile Savary, Ivan Balog and Leo Radzihovsky for inspiring discussions.
We acknowledge support from the French Agence Nationale de la Recherche by the Grant No.~ANR-17-CE30-0023 (DIRAC3D), ToRe IdexLyon breakthrough program, and ERC under the European Union's Horizon 2020 research and innovation program (project TRANSPORT No.853116). I.A.G. thanks the CNRS visiting researcher program and
the ENS de Lyon, where a part of this work was done, for hospitality.

\appendix

\section{Boundary conditions} \label{app:BC}

Let us consider a semi-infinite system with the boundary normal to the unit vector $\hat{n}_B$. Then the matrix $M$ should satisfy the following generalization of Eq.~\eqref{eq:BC-current}:
\begin{align}
\{M, \hat{n}_B \cdot \bm{\alpha} \} = \{ M, \tau_3 \otimes (\hat{n}_B \cdot \bm{\sigma}) \}
= 0.
\label{M-alpha-anticommutaator}
\end{align}
To find matrices $M$ that satisfy this relation, we follow the strategy of Ref.~\cite{Akhmerov-Boundary-2008} and start by expanding an arbitrary Hermitian matrix $M$ as
\begin{align}
M = \sum_{\mu,\nu = 0}^3 c_{\mu\nu} \tau_\mu \otimes \sigma_\nu
\end{align}
where the 16 coefficients $c_{\mu\nu}$ are real. Substitution of this form into Eq.~\eqref{M-alpha-anticommutaator} gives conditions on the coefficients $c_{\mu\nu}$ that can be written as
\begin{align}
c_{00} &= c_{30} = 0,
&
\mathbf{c}_0 \cdot \hat{n}_B &= \mathbf{c}_3 \cdot \hat{n}_B = 0,
\nonumber \\
\mathbf{c}_1 &= b_1 \hat{n}_B, & \mathbf{c}_2 &= b_2 \hat{n}_B,
\label{vector-constraints}
\end{align}
where we have introduced four 3-vectors
\begin{align}
\mathbf{c}_\mu = (c_{\mu 1}, c_{\mu 2}, c_{\mu 3}).
\end{align}
Combining the so-far unconstrained coefficients into two vectors in the $x$-$y$ plane,
\begin{align}
\mathbf{a} &= (c_{10}, c_{20}, 0),
&
\mathbf{b} &= (b_1, b_2, 0),
\end{align}
the matrix $M$ acquires the form
\begin{align}
M &= \tau_0 \otimes (\mathbf{c}_0 \cdot \bm{\sigma})
+ \tau_3 \otimes (\mathbf{c}_3 \cdot \bm{\sigma})
+ (\mathbf{a} \cdot \bm{\tau}) \otimes \sigma_0
\nonumber \\
& \quad + (\mathbf{b} \cdot \bm{\tau}) \otimes (\hat{n}_B \cdot \bm{\sigma}).
\label{M-vectors}
\end{align}

Next we substitute Eq.~\eqref{M-vectors} into the condition $M^2 = \mathbb{I}$. This gives several constraints. First of all, we get
\begin{align}
\mathbf{c}_0^2 + \mathbf{c}_3^2 + \mathbf{a}^2 + \mathbf{b}^2 &= 1,
& \mathbf{c}_0 \cdot \mathbf{c}_3 &= 0,
& \mathbf{a} \cdot \mathbf{b} &= 0.
\label{sum-of-squares}
\end{align}
Together with Eq.~\eqref{vector-constraints} this means that the three vectors $\mathbf{c}_0$, $\mathbf{c}_3$, and $\hat{n}_B$ are mutually orthogonal, and we can write
\begin{align}
\mathbf{c}_3 &= \lambda \hat{n}_B \times \mathbf{c}_0.
\end{align}
Another constraint is
\begin{align}
\mathbf{a} &= \lambda (\hat{z} \times \mathbf{b}).
\end{align}
Now the first equation in ~\eqref{sum-of-squares} becomes
\begin{align}
(1 + \lambda^2) (\mathbf{c}_0^2 + \mathbf{b}^2) = 1.
\end{align}
Solutions of this equation can be parametrized as
\begin{align}
\lambda &= \tan \theta,
\quad |\mathbf{c}_0| = \cos\theta \cos\gamma,
\quad |\mathbf{b}| = \cos\theta \sin\gamma,
\nonumber \\
\theta &\in [-\pi/2, \pi/2] \quad \gamma \in [0, \pi/2].
\end{align}
Let us use the angle $\phi \in [0, 2\pi)$ to characterize the direction of the vector $\mathbf{b}$ in the $x$-$y$ plane. Then
\begin{align}
\mathbf{b} &= \cos\theta \sin\gamma \, (\cos\phi, \sin\phi, 0),
\nonumber \\
\mathbf{a} &= \sin\theta \sin\gamma \, (-\sin\phi, \cos\phi, 0).
\end{align}
Finally, let us use the unit vector $\hat{n}_0$ in the direction of $\mathbf{c}_0$, so that
\begin{align}
\mathbf{c}_0 &= \cos\theta \cos\gamma \, \hat{n}_0,
&
\mathbf{c}_3 &= \sin\theta \cos\gamma \, (\hat{n}_B \times \hat{n}_0).
\end{align}

We end up with four angles: $\theta$, $\gamma$, $\phi$, and the angle determining the direction of $\hat{n}_0$ in the plane normal to $\hat{n}_B$. Then we can write the most general four-parameter family of $M$ matrices:
\begin{align}
M &= \cos\theta \cos\gamma \, \tau_0 \otimes (\hat{n}_0 \cdot \bm{\sigma})
\nonumber \\
& \quad + \sin\theta \cos\gamma \, \tau_3 \otimes (\hat{n}_B \times \hat{n}_0) \cdot \bm{\sigma}
\nonumber \\
& \quad - \sin\theta \sin\gamma \, (\sin\phi \, \tau_1 - \cos\phi \, \tau_2) \otimes \sigma_0
\nonumber \\
& \quad + \cos\theta \sin\gamma \, (\cos\phi \, \tau_1 + \sin\phi \, \tau_2)
\otimes (\hat{n}_B \cdot \bm{\sigma}).
\end{align}
Let us now simplify the matrix $M$ for the case $\hat{n}_B = \hat{z}$ that we are dealing with. In this case the unit vector $\hat{n}_0$ lies in the $x$-$y$ plane, and we can use another angle $\psi \in [0, 2\pi)$ to write
\begin{align}
\hat{n}_0 &= (\cos\psi, \sin\psi, 0),
\nonumber \\
\hat{n}_B \times \hat{n}_0 &= \hat{z} \times \hat{n}_0
= (-\sin\psi, \cos\psi, 0).
\end{align}
Then the matrix $M$ becomes
\begin{align}
M &= \cos\theta \cos\gamma \, \tau_0 \otimes (\cos\psi \, \sigma_1 + \sin\psi \, \sigma_2)
\nonumber \\
& \quad - \sin\theta \cos\gamma \, \tau_3 \otimes (\sin\psi \, \sigma_1 - \cos\psi \, \sigma_2)
\nonumber \\
& \quad - \sin\theta \sin\gamma \, (\sin\phi \, \tau_1 - \cos\phi \, \tau_2) \otimes \sigma_0
\nonumber \\
& \quad + \cos\theta \sin\gamma \, (\cos\phi \, \tau_1 + \sin\phi \, \tau_2)
\otimes \sigma_3,
\label{M-general}
\end{align}
which is exactly Eq.~\eqref{eq:M-general}.

%%%%%%%%%%%%%%%%%%%%%%%%%%%%%%% Surface spectrum in slab geometry %%%%%%%%%%%%%%%%%%%%%%%%

\begin{figure}[b!t]
\includegraphics[scale=0.47]{\PathToFigures 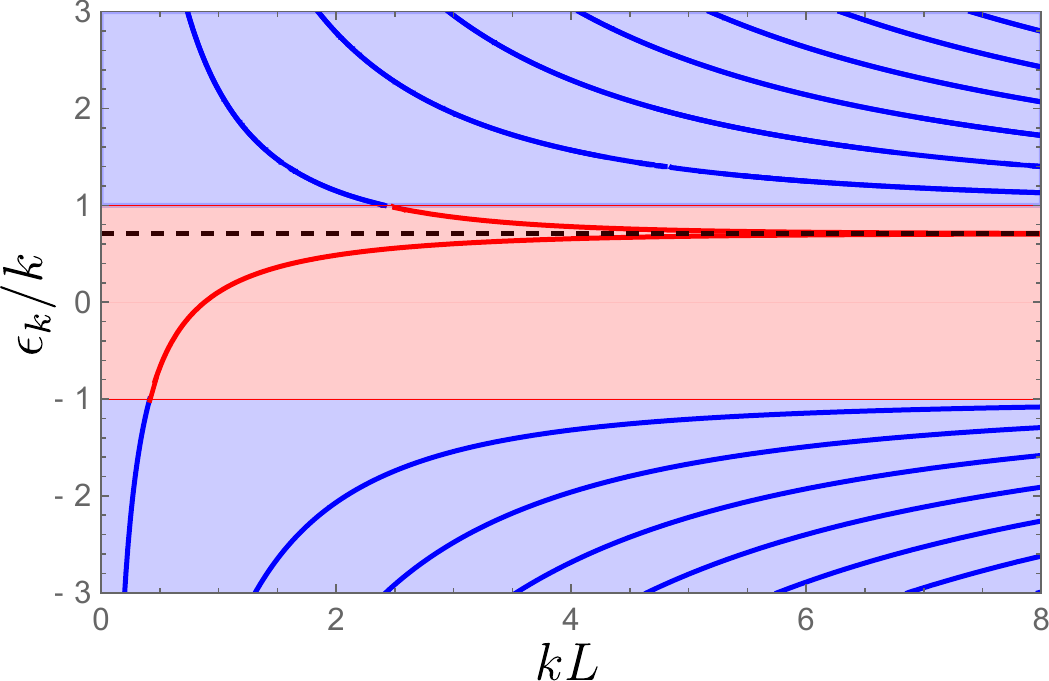}
\caption{Spectrum of the Dirac Hamiltonian~\eqref{eq:Ham0} in a slab geometry for symmetric BCs with $\theta=-\frac{\pi}4$. In the limit of $k L \gg 1$ the states with energies $|\epsilon_k|/k>1$ (blue lines) form the bulk Dirac bands while the two states (red lines) whose energy asymptotically approach
$\epsilon_k/k = 1/\sqrt{2}$ (dashed black line) are localized at the boundaries.}
\label{fig:surface-spectrum}
\end{figure}
%%%%%%%%%%%%%%%%%%%%%%%%%%%%%%%%%%%%%%%%%%%%%%%%%%%%%%%%%%%%%%%%%%%%

\section{Symmetry constraints on boundary conditions} \label{app:BC-symmetry}

The bulk Dirac Hamiltonian~\eqref{eq:Ham0} in a clean system is conformally invariant. The Euclidean conformal group in $d$ dimensions is generated by translations $P^\mu$, rotations $M^{\mu \nu}$, dilatation $D$, and special conformal generators $K^\mu$. These generate infinitesimal conformal transformations of Dirac spinors that can be written as
\begin{align*}
[P^\mu, \psi(x)] &= \partial^\mu \psi(x),
\nonumber \\
[M^{\mu \nu}, \psi(x)] &= (x^\nu \partial^\mu - x^\mu \partial^\nu + S^{\mu\nu}) \psi(x),
\nonumber \\
[D, \psi(x)] &= (x^\mu \partial_\mu + \Delta_\psi) \psi(x),
\nonumber \\
[K^\mu, \psi(x)] &= (2 x^\mu x^\nu \partial_\nu - x^2 \partial^\mu
+ 2 \Delta_\psi x^\mu - 2 x_\nu S^{\mu\nu}) \psi(x),
\end{align*}
where the generators $S^{\mu\nu}$ of internal rotations in the spinor space are
\begin{align}
S^{\mu\nu} &\equiv \frac{1}{4} [\gamma^\mu, \gamma^\nu] = \frac{1}{2} \gamma^\mu \gamma^\nu
\quad \text{ for } \quad \mu \neq \nu.
\end{align}
In our setting the role of the Euclidean gamma matrices is played by the \emph{alpha} matrices $\alpha_i$.

%%%%%%%%%%%%%%%%%%%%%%%%%%%%%%% Wave function in slab geometry %%%%%%%%%%%%%%%%%%%%%%%%

\begin{figure}[b!t]
%\centering
\includegraphics[scale=0.47]{\PathToFigures 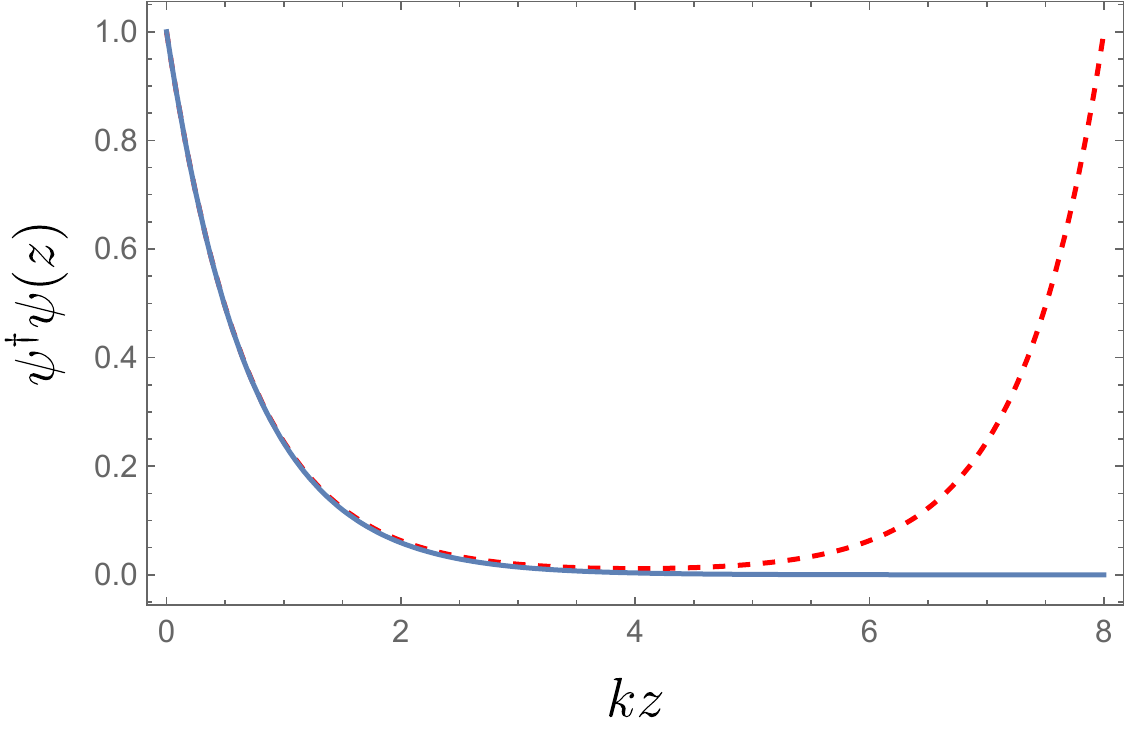}
\caption{Red dashed line: the lowest energy surface state  in a slab geometry  for symmetric BCs with $\theta=-\frac{\pi}4$ and $k L =8$. Blue solid line: the surface state in a semi-infinite geometry given by Eqs.~\eqref{eq:Dirac-solution00} and~\eqref{eq:Dirac-solution1} for the same parameters.}
\label{fig:wave-function}
\end{figure}
%%%%%%%%%%%%%%%%%%%%%%%%%%%%%%%%%%%%%%%%%%%%%%%%%%%%%%%%%%%%%%%%%%%%

When we consider a semi-infinite system with the boundary at $z = 0$, we can only ask whether the system is invariant under conformal transformations preserving the plane $z = 0$. Thus, we restrict the indices $\mu, \nu$ above to values 1 and 2. The boundary condition~\eqref{eq:BC-1} is conformally invariant if the matrix $M$ commutes with $D$, $P^1$, $P^2$, $M^{12}$, $K^1$, and $K^2$. The first three of these conditions are trivially satisfied for any constant (spatially-independent) matrix $M$. The last three conditions \emph{all} reduce to the requirement that $M$ commutes with $S^{12}$. In other words, rotational invariance in the $x$-$y$ plane turns out equivalent to conformal invariance. Since in our case $S^{12} \propto \alpha_1 \alpha_2 \propto \tau_0 \otimes \sigma_3$, we need to check the condition
\begin{align}
[M, \tau_0 \otimes \sigma_3] &= 0.
\label{M-conf-inv-condition}
\end{align}

It is easy to see that the last two terms in a general matrix~\eqref{eq:M-general} satisfy Eq.~\eqref{M-conf-inv-condition} for any value of the parametrizing angles, but the first two terms do not. These symmetry-violating terms both vanish when $\gamma = \pi/2$, in which case we get the two-parameter family~\eqref{eq:M-conformal} (upon making an additional trivial shift $\phi \to \phi + \pi$ to flip the overall sign):
\begin{align}
M_\mathrm{3D conf} &= \sin\theta \, (\sin\phi \, \tau_1 - \cos\phi \, \tau_2) \otimes \sigma_0 \nonumber \\
& \quad - \cos\theta \, (\cos\phi \, \tau_1 + \sin\phi \, \tau_2) \otimes \sigma_3.
\end{align}

%%%%%%%%%%%%%%%%%%%%%%%%%%%Figure 1 %%%%%%%%%%%%%%%%%%%%%%%%%%%%%%%%%%%%%%%%%%%%%%%
\begin{figure}[t]
\includegraphics[width=75mm]{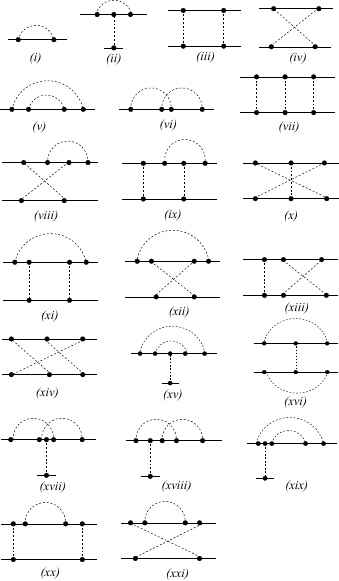}
\caption{Diagrams contributing to the one- and two-particle vertex functions to two-loop order.}
 \label{fig:two-loop-diagrams}
\end{figure}
%%%%%%%%%%%%%%%%%%%%%%%%%%%%%%%%%%%%%%%%%%%%%%%%%%%%%%%%%%%%%%%%%%%%%%%%%%

%%%%%%%%%%%%%%%%%%%%%%%%%%%%%Table 1%%%%%%%%%%%%%%%%%%%%%%%%%%%%%%%%%%%%%%%%%%%%%
\begin{table}[t]
\begin{tabular}{|l|l|} \hline
Diagram & Expression \\
\hline
(i) & $\dfrac{K_d}2 \mathring{\omega}^{\varepsilon} \mathring{\Delta} \dfrac{2 i \mathring{\omega}}{\varepsilon}$ \\[1mm] \hline
(ii) & $- \dfrac{K_d}2 \mathring{\omega}^{\varepsilon} \mathring{\Delta}^2  \left(\dfrac4{\varepsilon}+6\right)$ \\[1mm] \hline
(iii)+ (iv)  & $0$ \\[1mm] \hline
(v) & $\dfrac{K_d^2}4 \mathring{\omega}^{2\varepsilon} \mathring{\Delta}^2 \left( - \dfrac{4 i \mathring{\omega}}{\varepsilon^2}-\dfrac{4 i \mathring{\omega}}{\varepsilon} \right)$ \\[1mm] \hline
(vi) & $ \dfrac{K_d^2}4 \mathring{\omega}^{2\varepsilon} \mathring{\Delta}^2 \left(\dfrac{\vec{\alpha} \cdot \vec{p} }{\varepsilon}- \dfrac{2 i \mathring{\omega}}{\varepsilon^2}  \right) $ \\[1mm] \hline
(vii)+(x)  & $\dfrac{K_d^2}4 \mathring{\omega}^{2\varepsilon} \mathring{\Delta}^3 \left(\dfrac{4}{\varepsilon^2}+\dfrac{2}{\varepsilon}\right)$ \\[1mm] \hline
(viii)+(ix) & $\dfrac{K_d^2}4 \mathring{\omega}^{2\varepsilon} \mathring{\Delta}^3 \left( \dfrac{16}{\varepsilon} \right)$ \\[1mm] \hline
(xi)+(xii) & $0$ \\[1mm] \hline
(xiii) + (xiv)  & $ \dfrac{K_d^2}4 \mathring{\omega}^{2\varepsilon} \mathring{\Delta}^3 \left( - \dfrac{4}{\varepsilon^2}-\dfrac{4}{\varepsilon} \right) $ \\[1mm] \hline
(xv) & $\dfrac{K_d^2}4 \mathring{\omega}^{2\varepsilon} \mathring{\Delta}^3 \left(\dfrac{8}{\varepsilon^2} + \dfrac{16}{\varepsilon} \right)$\\[1mm] \hline
(xvi)  & $ \dfrac{K_d^2}4 \mathring{\omega}^{2\varepsilon} \mathring{\Delta}^3 \left( \dfrac{4}{\varepsilon^2}+\dfrac{8}{\varepsilon}\right)$\\[1mm] \hline
(xvii)  & $\dfrac{K_d^2}4 \mathring{\omega}^{2\varepsilon} \mathring{\Delta}^3 \left( -\dfrac{4}{\varepsilon^2}-\dfrac{8}{\varepsilon}\right)$\\[1mm] \hline
(xviii)  & $\dfrac{K_d^2}4 \mathring{\omega}^{2\varepsilon} \mathring{\Delta}^3 \left( \dfrac{8}{\varepsilon^2}+\dfrac{16}{\varepsilon} \right)$\\[1mm] \hline
(xix)  & $ \dfrac{K_d^2}4 \mathring{\omega}^{2\varepsilon} \mathring{\Delta}^3 \left(\dfrac{8}{\varepsilon} \right)$ \\[1mm] \hline
(xx)+ (xxi)  & $0$ \\[1mm] \hline
\end{tabular}
\caption{Integrals corresponding to the diagrams shown in Fig.~\ref{fig:two-loop-diagrams} computed using dimensional regularization with
their combinatorial factors.}
\label{tab:diag}
\end{table}
%%%%%%%%%%%%%%%%%%%%%%%%%%%%%%%%%%%%%%%%%%%%%%%%%%%%%%%%%%%%%%%%%%%%%%%%%%

If we also require invariance under time reversal $\mathcal{T} = \tau_0 \otimes \sigma_2 \mathcal{K}$, we need to satisfy $[\mathcal{T}, M] = 0$ or
\begin{align}
M \tau_0 \otimes \sigma_2 &= \tau_0 \otimes \sigma_2 M^*.
\label{T-inv-M}
\end{align}
It is easy to check that the general matrices~\eqref{eq:M-general} satisfy Eq.~\eqref{T-inv-M} only when $\gamma = \pi/2$ and $\phi = \pm \pi/2$, which gives two families $\pm M_\theta$ where
\begin{align}
%M_{T\pm} &= \pm M_\theta,
%&
M_\theta &= \sin\theta \, \tau_1 \otimes \sigma_0
- \cos\theta \, \tau_2 \otimes \sigma_3.
\label{M-T-family}
\end{align}
These matrices are contained within the conformally-invariant family~\eqref{eq:M-conformal}, and are
unitary-equivalent to any matrix $M_\text{3D conf}$: $M_\text{3D conf} = \pm U M_\theta U^{-1}$, where
\begin{align}
U &= e^{i \tilde{\phi}_\pm \tau_3} \otimes \sigma_0
= (\cos\tilde{\phi}_\pm \, \tau_0 + i \sin\tilde{\phi}_\pm \, \tau_3) \otimes \sigma_0,
\nonumber \\
\tilde{\phi} &= \pm \frac{\pi}{4} - \frac{\phi}{2}.
\end{align}
These unitary transformations amount to rotations of $\tau$ matrices in the $x$-$y$ plane and, therefore, do not change the matrices $\alpha_i$ and the massless Dirac Hamiltonian~\eqref{eq:Ham0}.

Finally, we may impose symmetry under charge conjugation $\mathcal{C} = \tau_2 \otimes \sigma_2 \mathcal{K} $: $[\mathcal{C}, M] = 0$ or
\begin{align}
M \tau_2 \otimes \sigma_2 &= \tau_2 \otimes \sigma_2 M^*.
\label{C-inv-M-2}
\end{align}
In the general case this gives two families
\begin{align}
M_{C1} &= \tau_3 \otimes (\sin\psi \, \sigma_1 - \cos\psi \, \sigma_2),
\nonumber \\
M_{C2} &= (\cos\phi \, \tau_1 + \sin\phi \, \tau_2) \otimes \sigma_3.
\label{M-C-family-2}
\end{align}
Only the second of these two families is contained in the conformally invariant family~\eqref{eq:M-conformal} when $\theta = 0$. As before, this family can be unitarily rotated to $M_{\theta = 0}$.

\section{Dirac surface states in a slab geometry}
\label{appendix:slab}

Here we consider the time-independent, massless Dirac equation $\hat{H}_0\psi = \epsilon \psi$ with Hamiltonian~\eqref{eq:Ham0} in a slab geometry where the system has two flat boundaries at $z=0$ and $z=L$. At each boundary we impose a BC of the form~\eqref{BC} with $\theta = \theta_1$ at $z=0$ and $\theta = \theta_2$ at $z=L$. We can again use the translational invariance along the two boundaries to perform the Fourier transform in the directions parallel to them, $ \vec{r} = (x,y) \to \vec{k} = (k_1,k_2)$, and look for solutions in the form
\begin{align}
\psi(\mathbf{r}) &= \psi_{\vec{k}}(z) e^{i \vec{k}\cdot \vec{r}}.
\end{align}
Similarly to Eqs.~\eqref{eq:Dirac1} and~\eqref{BC}, we now have to solve
\begin{align}
(-i \, \alpha_3 \partial_3 + \vec{\alpha} \cdot \vec{k})\psi_{\vec{k}}(z)
&= \epsilon_k \psi_{\vec{k}}(z),
\label{eq:Dirac1-slab}
\\
M_{\theta_1} \psi_{\vec{k}}(0) &= \psi_{\vec{k}}(0), \\
M_{\theta_2} \psi_{\vec{k}}(L) &= \psi_{\vec{k}}(L).
\end{align}

For simplicity, we consider the symmetric case with identical boundaries. Then the symmetry $z \to -z$, $\theta \to -\theta$ implies that the BC angles at the two boundaries are related as $\theta_1 = -\theta_2 = \theta$. In this case the spectrum $\epsilon_k$ is determined by the equation
\begin{align}
\tanh\Big(L \sqrt{k^2-\epsilon_k^2}\Big)(\epsilon_k  \cos \theta \pm k )= \sqrt{k^2-\epsilon_k^2}\sin \theta,
\label{eq-surf-slab-1}
\end{align}
where the upper sign corresponds to positive $\theta \in (0,\pi/2]$ and the lower sign to negative $\theta \in [-\pi/2, 0)$. This spectrum is shown in Fig.~\ref{fig:surface-spectrum} as a function of $L$ for $\theta=-\frac{\pi}4$. For large slab thickness ($k L |\sin\theta|\gg 1$), Eq.~\eqref{eq-surf-slab-1} has two solutions with $|\epsilon_k|<k$  given by
\begin{align}
\epsilon_k &\approx \epsilon_{0k} \pm 2 k  \sin^2\theta \exp(-k L |\sin\theta|),
%&
%k L |\sin\theta| &\gg 1,
\label{eq-surf-slab-2}
\end{align}
which are localized at the boundaries. In the limit of $k L |\sin\theta| \to \infty$ their energies asymptotically approach $\epsilon_{0k}=\mp k\cos\theta$, which is the energy of the surface state in the semi-infinite geometry; see Eq.~\eqref{eq:esp-k}.

For $\theta = 0$  there are no surface states, either in the slab geometry or in a semi-infinite system. Otherwise ($\theta \neq 0$), the surface states on both boundaries, being infinitely separated ($L\to \infty$),  would have exactly  the same energy due to the choice of the BCs. For finite $L$ the degeneracy is lifted, similar to the problem of a symmetric double well in nonrelativistic quantum mechanics. The density $\psi^\dagger \psi(z)$ for the lower energy surface state in the slab geometry and the density for the wave function in the semi-infinite geometry, normalized to $\psi^{\dagger}\psi(0)=1$, are shown in Fig.~\ref{fig:wave-function}.

The solutions with  $|\epsilon_k|>k$ describe the bulk Dirac bands with energies $\epsilon_k \approx \pm \sqrt{k^2+(n\pi+\theta)^2/L^2}$, $n \in \mathbb{N}$  for $|\epsilon_k| \gg k$. Here we considered the simplest case of a single Dirac cone. Surface states in a slab geometry for Weyl semimetal with several cones  were discussed in Ref.~\cite{Benito-Matias-Molina:2019}.

\section{Vertex functions in the bulk} \label{Appendix-bulk}

The one- and two-particle vertex functions in the bulk are given to two-loop order by the diagrams shown in Fig.~\ref{fig:two-loop-diagrams}. The corresponding integrals are computed using dimensional regularization and are given in Table~\ref{tab:diag}.

\section{Calculation of the one-loop integrals in real space}
\label{Appendix-integrals}

The integral
\begin{align}
I = \int d^{d-1}r \int\limits_0^{\infty} dz  \frac{z^{2-d}}{[(z^2+(\vec{r}-\vec{r}_1)^2)(z^2+(\vec{r}-\vec{r}_2)^2)]^{d/2}}
\end{align}
that appears in Eq.~\eqref{eq-dG-1} can be computed using the Feynman parametrization. First we write
\begin{widetext}
\begin{center}
\begin{align}
\frac{1}{[z^2 + (\vec{r} - \vec{r}_1)^2]^{d/2} [z^2 + (\vec{r} - \vec{r}_2)^2]^{d/2}}
&= \frac{\Gamma(d)}{\Gamma^2(d/2)}\int\limits_0^{1} d\beta
\frac{\beta^{d/2-1} (1-\beta)^{d/2-1}}{\big(\beta[z^2 + (\vec{r} - \vec{r}_1)^2] + (1-\beta)[z^2 + (\vec{r} -\vec{r}_2)^2]\big)^d}.
\label{eq:Feynman}
\end{align}
\end{center}
This expression will be integrated over $\vec{r}$. Therefore we can make a linear change of variables to remove the terms linear in $\vec{r}$ in the denominator:
\begin{align}
\vec{r} \to \vec{r}+\beta \vec{r}_1 +(1-\beta)\vec{r}_2.
\label{linear-shift}
\end{align}
This gives
\begin{align}
I &= \frac{\Gamma(d)}{\Gamma^2\big(\frac{d}{2}\big)} \int\limits_0^{1} \! d\beta \, \beta^{d/2-1} (1-\beta)^{d/2-1} \int\limits_0^{\infty} \! dz \, z^{2-d} \int \frac{d^{d-1}r}{[r^2 + z^2 + \beta(1-\beta) r_{12}^2]^d}.
\end{align}

The subsequent integrals over $\vec{r}$ and $z$ converge for $1 < d < 3$ and are easily reduced to the beta function with the result
\begin{align}
\int \frac{d^{d-1}r}{[r^2 + z^2 + \beta(1-\beta) r_{12}^2]^d}
&= \pi^{(d-1)/2} \frac{\Gamma\big(\frac{d+1}2\big)}{\Gamma(d)}
\frac{1}{[z^2 + \beta(1-\beta) r_{12}^2]^{(d+1)/2}},
\nonumber \\
\int\limits_0^{\infty} \! dz \frac{z^{2-d}}{[z^2 + \beta(1-\beta) r_{12}^2]^{(d+1)/2}}
&= \frac{ \Gamma \big(\frac{3-d}{2}\big) \Gamma (d-1)}{2
\Gamma \big(\frac{d+1}{2}\big)} \frac{1}{[\beta(1-\beta) r_{12}^2]^{d-1}}.
\end{align}
\end{widetext}
Finally we do the $\beta$-integral which converges for $d < 2$:
\begin{align}
\int\limits_0^{1} \! d\beta \, [\beta(1-\beta)]^{-d/2}
= \frac{\Gamma^2\big(1-\frac{d}{2}\big)}{\Gamma (2-d)}.
\end{align}
Combining all factors and using the Legendre duplication formula for the gamma function
\begin{align}
\Gamma(2-d) &= \pi^{-1/2} 2^{1-d} \Gamma\big(1 - \tfrac{d}{2}\big) \Gamma\big(\tfrac{3-d}{2}\big),
\end{align}
we get for $1 < d < 2$
\begin{align}
I &= 2^{d-2} \pi^{d/2} \frac{\Gamma (d-1) \Gamma\big(1 - \tfrac{d}{2}\big)}{\Gamma^2\big(\frac{d}{2}\big)}
\frac{1}{r_{12}^{2d - 2}}.
\end{align}

To derive Eq.~\eqref{eq:delta2G} from Eq.~\eqref{eq-dG-2} we use the Feynman parametrization~\eqref{eq:Feynman} and then change variables $\vec{r} \to \vec{r}+\beta \vec{r}_1 +(1-\beta)\vec{r}_2 $. Omitting the linear in $\vec{r}$ terms which integrate out to zero we arrive at Eq.~\eqref{eq:delta2G}.

\vspace{5mm}

%%% BIBLIO %%%

\bibliographystyle{apsrev4-2}
%\bibliography{biblioSurface}

%apsrev4-2.bst 2019-01-14 (MD) hand-edited version of apsrev4-1.bst
%Control: key (0)
%Control: author (72) initials jnrlst
%Control: editor formatted (1) identically to author
%Control: production of article title (-1) disabled
%Control: page (0) single
%Control: year (1) truncated
%Control: production of eprint (0) enabled
%

\end{document}